\documentclass[fleqn,usenatbib]{mnras}

\usepackage{newtxtext,newtxmath}

\usepackage[T1]{fontenc}

\DeclareRobustCommand{\VAN}[3]{#2}
\let\VANthebibliography\thebibliography
\def\thebibliography{\DeclareRobustCommand{\VAN}[3]{##3}\VANthebibliography}

\usepackage{xcolor}

\usepackage{graphicx}	
\usepackage{amsmath}	
\usepackage{array, booktabs}

\title[Shadows cast by precessing annuli in HD~139614]{Shadows cast by precessing annuli as evidence for an inclined planet in HD~139614}

\author[Katie L. Milsom et al.]{
Katie L. Milsom,$^{1}$
Tim J. Harries,$^{1}$
Evan A. Rich,$^{2}$
John D. Monnier,$^{3}$
Anna S. E. Laws,$^{4,\, 5}$
\newauthor{Benjamin R. Setterholm$^{3,\,6}$}
\\\\
$^{1}$Department of Physics and Astronomy, University of Exeter, Stocker Rd, Exeter EX4 4QL, UK\\
$^{2}$ University of Nebraska-Lincoln, 1400 R St., Lincoln, NE, 68588, USA\\
$^{3}$Department of Astronomy, University of Michigan, Ann Arbor, MI 48109, USA\\
$^{4}$University of Exeter Medical School, Exeter, UK\\
$^{5}$NIHR Applied Research Collaboration South West (ARC South West), Exeter, UK\\
$^{6}$ Max-Planck-Institut für Astronomie, Heidelberg, Germany 
}

\date{Accepted 2026 September 28. Received 2026 September 23; in original form 2026 June 10}

\pubyear{2025}

\begin{document}
\label{firstpage}
\pagerange{\pageref{firstpage}--\pageref{lastpage}}
\maketitle

\begin{abstract}
We analyse scattered-light images at multiple epochs of the protoplanetary disc HD~139614 taken with GPI and SPHERE instruments. We show that the azimuthal distribution of the surface brightness of the inner bright ring ($13-21\,$au) varies with time (likely caused by a structure in the inner-disc at $\sim1\,$au) and constrain the timescale of this variation to be $<11\,$months. We show that inner arc ($31-44\,$au) is likely a spiral and tightly wound, measuring a pitch angle of $6.1\pm1.2^\circ$. Additionally, we construct a physical model to explain the broad shadow across $\sim2/3$ of the outer disc and the azimuthally asymmetric bright ring (which has a shadow not aligned with the broad shadow). We model the disc using 1D warp equations with a fast radiative transfer code to produce scattered-light images. We setup the disc as three annuli separated by gaps with an inclined perturbing planet between the inner two annuli. The planet causes the inner annuli to tilt and precess independently casting broad shadows across the outer (un-tilted) annulus. These shadows combine to produce a shadow across $\sim2/3$ of the outer disc and an asymmetric bright ring which matches the observations of HD~139614. The precession of the annuli is periodic (so the model matches the observations multiple times) with long precession time-scales ($8.4\times10^3$ and $7.2\times10^4\,$years for the inner and middle annulus respectively) so the shadows are effectively static over accessible observational time-scales.
\end{abstract}

\begin{keywords}
planet-disc interactions -- protoplanetary discs -- stars:protostars -- techniques: polarimetric -- methods: numerical -- radiative transfer
\end{keywords}



\section{Introduction}
\label{sec: introduction}
Protoplanetary discs orbit protostars and are the birthplace of planetary systems (see review by \citealt{andrews_2020}). High-resolution instruments have revealed large-scale structures in these discs, including cavities, rings, gaps, arcs, and spirals (e.g. \citealt{garufi_2018, rich_2022}), and in some cases these are thought to be directly linked to the formation and presence of planets (see overview by \citealt{espaillat_2014}). It is clear that understanding the morphology and formation of these structures is key to our understanding of planet formation itself.

Protoplanetary discs can be directly imaged using light scattered from the surface of the disc (eg. \citealt{fischer_1995, fischer_1996}). Small, $\rm \mu m$-sized, grains are dynamically coupled to the gas in the disc and so trace its surface. Starlight scattered off these grains is polarised (while the stellar light is un-polarised) so observations of the scattered, polarised light allows us to image the surface of the disc. Polarimetric Differential Imaging (PDI) can be used to improve the contrast between the disc and the host star by isolating the linearly polarised scattered-light and reducing the stellar PSF (as outlined in \citealt{kuhn_2001}).

Scattered-light images are particularly sensitive to departures in the morphology of the disc from a smooth surface. Therefore, perturbations in the inner disc, or features such as a misaligned inner disc or a warped inner disc, can cast shadows across the outer disc which are visible in the scattered-light images (eg. \citealt{pinilla_2015,stolker_2016}). Narrow shadows can be cast by a misaligned inner disc with a fairly large angle of inclination compared to the outer disc (eg. two narrow shadows \citealt{marino_2015, benisty_2017, casassus_2018}). In contrast, broad shadows (across up to half of the outer disc) are cast when the misaligned inner disc has a low angle of misalignment compared to the outer disc (eg. \citealt{benisty_2018, bohn_2019}). 

Shadows can also vary azimuthally. For example, TW Hya exhibits time-variable shadows, which \citet{debes_2017} attribute to an inclined precessing inner disc ($<1\,$au), potentially driven by a planet on a misaligned orbit. In fact \citet{nealon_2018, nealon_2019} use hydrodynamic simulations to demonstrate that a planet on an inclined orbit can open a gap dividing the disc into an outer region, and an inner, tilted, precessing disc which casts a moving shadow (see also \citealt{zhu_2019} for inner disc precession). More recent observations of TW Hya, presented in \citet{debes_2023}, show that the shadow continues to move, but evolves into two separate shadows. They explain these shadows with two manually misaligned, precessing disc components each casting a shadow which can overlap to appear as one shadow (as in the earlier observations).

Misaligned discs can also be induced by binary companions (eg. \citealt{ballabio_2021}). If the stellar mass ratio is sufficiently large and the disc is significantly misaligned, with respect to the binary orbit, ($\gtrsim40^\circ$ for a mass ratio $\mu>0.1$ and an eccentricity of $e=0$ \citealt{young_2023}), disc tearing may occur, fragmenting the disc into rings which orbit the star on misaligned planes \citep{facchini_2013, nixon_2013, facchini_2018}. However, since this process generally requires a high degree of misalignment (except for very eccentric binaries, $e>0.8$ and $\mu=0.5$, where misalignments of $\gtrsim25^\circ$ are enough \citealt{young_2023}), the shadows cast are likely to be narrow and diametrically opposite rather than broad shadows. For example \citet{kraus_2020} attribute the shadows observed in GW Ori to misaligned annuli produced from disc tearing (although, \citealt{smallwood_2021,smallwood_2025} suggest that the disc may be separated into annuli by massive planets).

Although misaligned discs may be signposts for planets, other causes of the misalignment could be due to the interaction between the disc and the stellar magnetosphere \citep{bouvier_1999} or the misalignment could occur as the disc formed \citep{bate_2018}. To confirm whether there is a misalignment, the inner disc can be analysed with interferometry (eg. \citealt{bohn_2022}) or by looking at disc kinematics (eg. \citealt{loomis_2017, perez_2018, young_2022}).

In this paper we focus on scattered-light observations of the protoplanetary disc around HD~139614, which shows both broad shadows and indirect evidence for a massive planet.

\subsection{HD~139614}
\label{sec: intro_hd139614}
HD~139614 is an A7-type Herbig star located in Sco-Cen OB in the Upper Centarus-Lupus subgroup (with a $99\%$ likeliness from the BANYAN $\Sigma$ association finder: \citealt{gagne_2018}). The star is located $\sim$133.6\,pc away \citep{gaia_2020} and with an age of $\sim 15.6$\,Myr \citep{fairlamb_2015}, is relatively old for a Herbig star (see Table \ref{tab:stellar_params} for the stellar parameters).

From the object's SED it has been determined that the disc is a flared (Group I) disc and is a pre-transitional disc, as there is a dip in part of the mid-infrared  \citep{matter_2014}. The disc is almost face-on with an inclination of $17.6^\circ$ (where $0^\circ$ is face-on and $90^\circ$ is edge-on) and a position angle (PA) of $276.5^\circ$ \citep{muro_arena_2020} (where PA is the angle between North and the semi-major axis, measured in the North to East direction). The disc properties are summarised in Table \ref{tab:disc_params}.

The inner disc has been observed by \citet{matter_2014, matter_2016} using interferometry. Their analysis of the interferometric data with geometrical and radiative transfer modelling suggest that the disc has an inner and outer region with a gap between them in the dust distribution from $\sim2.5-6\,$au \citep{matter_2016}. The gap is potentially caused by a $\sim3\rm \,M_{jup}$ mass planet at $\sim4\,$au clearing the disc material, and the plausibility of this hypothesis was demonstrated by \citet{matter_2016} using hydrodynamical simulations. The presence of this planet was supported by \citet{carmona_2017} by analysis of CO ro-vibrational emission lines to estimate the gas distribution in the disc; factors such as a gas density drop between $1-6$\,au (despite the fact that the star is still accreting) suggest the presence of a $<2\rm \,M_{Jup}$ planet at $\sim4$\,au.

The surface of the disc has been observed in scattered-light (eg. \citealt{laws_2020, muro_arena_2020}) and the dust continuum has been observed with ALMA (see \citet{stapper_2022} and references therein). However the resolution of the ALMA data is $0.74''$ and there are as yet no detailed ALMA images of the disc in which substructures can be seen. 

Various features of the disc can be seen in the scattered-light images, including two arcs which could possibly be part of spiral arms (as suggested by \citealt{laws_2020}). The scattered-light observations show a broad asymmetric shadow across $\sim2/3$ of the outer disc and an azimuthally asymmetric bright ring further in. 

The shadow cannot be explained with just one misaligned region as it spans more than half the outer disc. Therefore, \citet{muro_arena_2020} modelled the disc using a geometric configuration consisting of a series of inclined annuli. Although arbitrary in a physical sense, their model accurately reproduces the bright ring, arcs, and broad shadow seen in observations of HD~139614. In this paper we aim create a physically-motivated model of the disc using a 1D warp model combined with a radiative transfer code to model scattered-light observations. We aim to reproduce features such as the broad shadow and inner bright ring seen in the scattered-light images.  

This paper is split into two parts: observations and modelling. In section \ref{sec: observational Data} we present the observational data for HD~139614, our analysis of the scattered-light images (Section \ref{sec: analysis_obs}), and a discussion of our findings (Section \ref{sec: discussion_obs}). In Section \ref{sec: modelling} we introduce our 1D warp model (Section \ref{sec: 1d_disc_model}), detail the set-up of our model (Section \ref{sec: model setup}), and outline our fast radiative transfer code (Section \ref{sec: radiative_transfer}). We present our model results (Section \ref{sec: model_results}), compare our model to observations, and outline the caveats of our model in the discussion (Section \ref{sec: model_discussion}). Finally we present our conclusions in Section \ref{sec: conclusion}.

\begin{table}
	\centering
	\caption{Table of stellar parameters for HD139614.}
	\label{tab:stellar_params}
	\begin{tabular}{lcr} 
        \hline
        Parameter               & Value                      & Reference \\
		\hline
        GAIA Distance [pc]      & $133.6\pm0.5\phantom{00}$             & {\citet{gaia_2020}} \\
        SpTyp                   & A7V                                   & {\cite{meeus_2001}} \\
        $M_* [M_{\odot}]$       & $1.481\pm0.074$                       & {\cite{vioque_2018}} \\
        Age [Myr]               & $15.64\pm4.29\phantom{0}$             & {\cite{fairlamb_2015}} \\
        $T_{\rm eff}[K]$        & $\phantom{.}7750\pm250\phantom{.0}$   & {\cite{vioque_2018}} \\
        $R_* [R_{\odot}]$       & $\phantom{00}1.4\pm0.2\phantom{00}$   & {\cite{fairlamb_2015}} \\  
        $L_* [L_{\odot}]$       & $5.929\substack{+0.454\phantom{0} \\ -0.135\phantom{0}}$    & {\cite{vioque_2018}} \\ [3pt]
        $\log(\dot{M}_{\rm acc} [M_\odot yr^{-1}])$ & $-7.63\substack{+0.20\phantom{00.} \\ -0.30\phantom{00.}}$ & {\cite{fairlamb_2015}} \\ [3pt]
		\hline
    \end{tabular}
\end{table}

\begin{table}
	\centering
	\caption{Table of disc parameters for HD139614.}
	\label{tab:disc_params}
	\begin{tabular}{lcr} 
        \hline
        Parameter                           & Value         & Reference \\
		\hline
         PA [$^\circ$]                      & $276.5\pm3.1\phantom{00}$              & {\citet{muro_arena_2020}} \\
         $i$ $[^\circ]$                     & $\phantom{0}17.6\pm3.1\phantom{00}$    & {\citet{muro_arena_2020}} \\
         $M_{\rm dust}[M_{\oplus}]$         & $\phantom{0}41.7\pm4.3\phantom{00}$    & {\citet{stapper_2022}} \\
        Inner disc: $r_{\rm outer}$ [au]   & $\phantom{0}2.55\pm0.13\phantom{0}$              & {\citet{matter_2016}}\\
         Outer disc: $r_{\rm inner}$ [au]   & $\phantom{00}5.7\substack{+0.3\phantom{000} \\ -0.2\phantom{000}}$  & {\citet{matter_2016}}\\ [3pt]
		\hline
    \end{tabular}
\end{table}

\section{Observational Data}
\label{sec: observational Data}

\begin{table}
	\centering
	\caption{Observational log for HD139614. References: \protect\cite{rich_2022} for GPI; \protect\citet{muro_arena_2020} for SPHERE.}
	\label{tab:observations}
	\begin{tabular}{lcccc} 
		\hline
		UT & Instrument & Band & $\#$ of Frames & Exposure time [s]\\
		\hline
		31-03-2016 & SPHERE & J & 56 & 64\\
  		06-04-2017 & GPI & J & 64 & 58.2\\
    	16-05-2017 & SPHERE & H & 32 & 64\\
     	08-06-2018 & GPI & H & 28 & 58.2\\
      	21-06-2018 & SPHERE & J & 32 & 64\\
       	13-05-2019 & GPI & J & 36 & 58.2\\
		\hline
	\end{tabular}
\end{table}

\begin{figure*}
	\includegraphics[width=\textwidth]{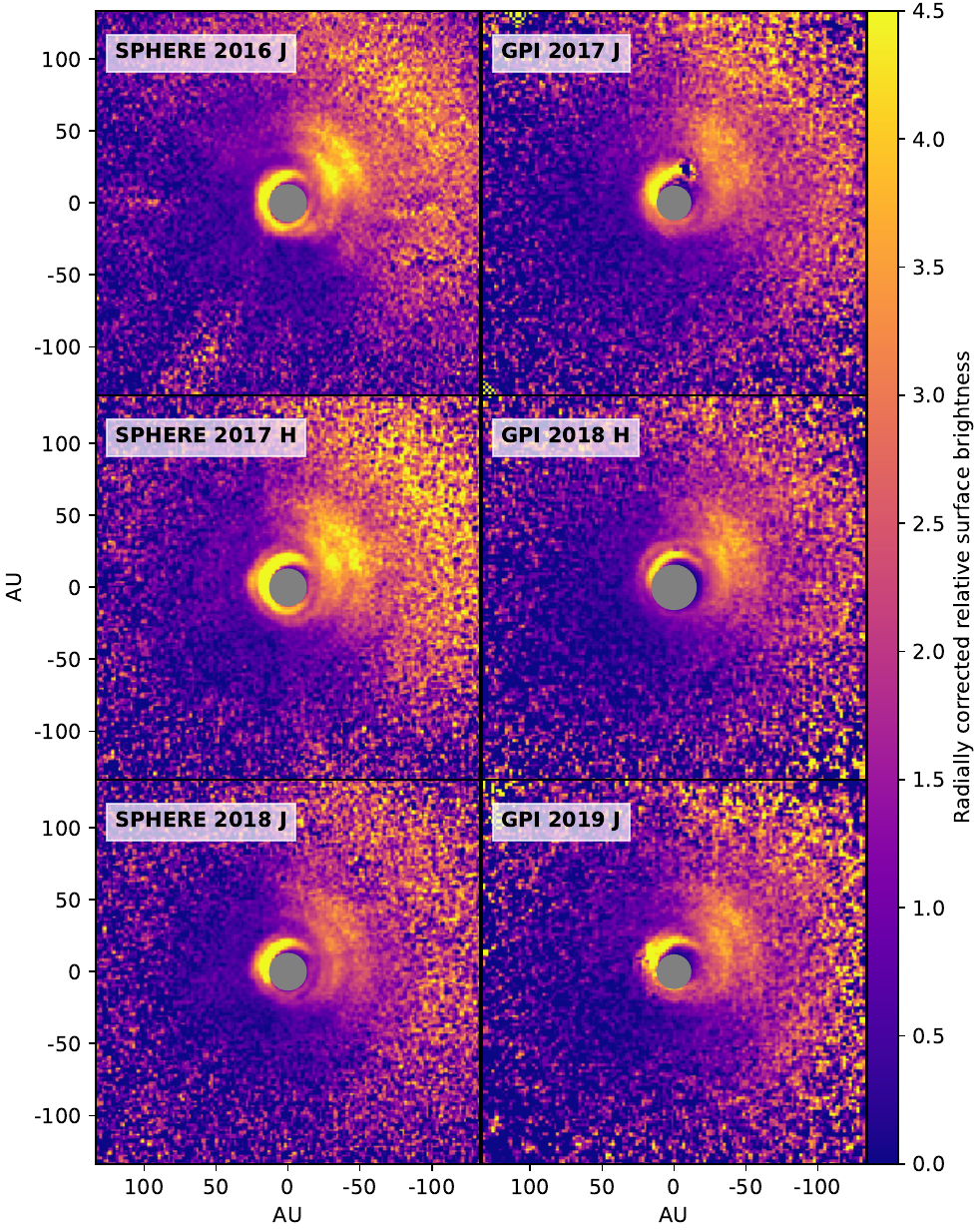}
    \caption{De-projected $Q_{\phi}$ images of SPHERE (left) and GPI (right) data for HD139614. The data has been $R^2$ scaled and normalised. The labels show the instrument, epoch, and wavelength band for each image. The coronagraphic spot is marked by a grey circle and North is up and East is left for all images. The 2017 and 2018 epochs have data from both GPI and SPHERE.}
    \label{fig: deprojected_data_full_plot}
\end{figure*}

\begin{figure*}
	\includegraphics[width=\textwidth]{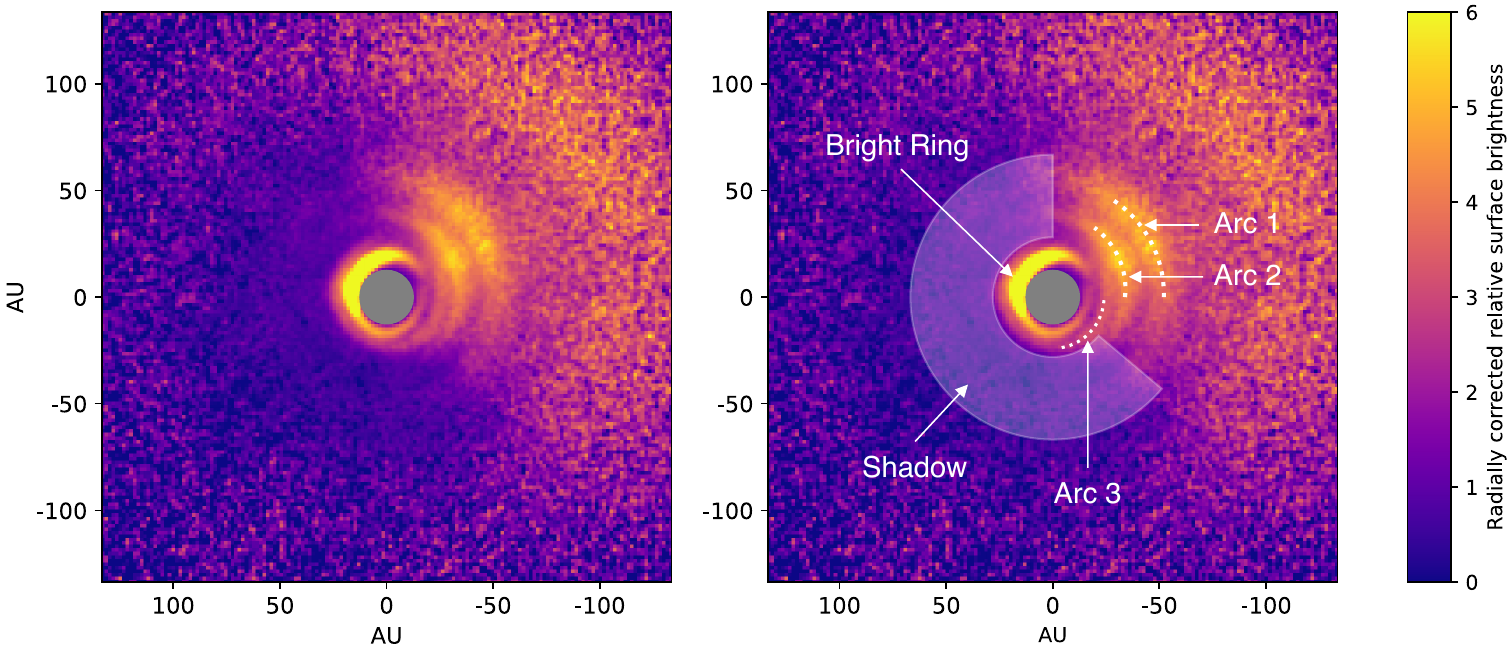}
    \caption{Left: De-projected, stacked $Q_{\phi}$ image of the SPHERE data. All epochs for SPHERE were stacked into one image, then de-projected, $R^2$ scaled and normalised. The coronagraphic spot is marked by a grey circle. North is up, East is left. Right: The same image as on the left, but labelled with the key features of the disc: an inner bright ring, three arcs, and a broad shadow across the outer disc. }
    \label{fig: stacked_sphere_qphi}
\end{figure*}

The scattered-light observations were obtained using the Gemini Planet Imager (GPI) \citep{macintosh_2014, poyneer_2014,larkin_2014} and Spectro-polarimetric High-contrast Exoplanet REsearch (SPHERE) \citep{beuzit_2019} instruments in Polarimetric Differential Imaging (PDI) mode \citep{perrin_2015}. The observation log is summarised in Table \ref{tab:observations}. The data were taken in the near-infrared, using waveband filters $J$ and $H$, with a coronagraph to block out the light from the host star. Adaptive optics were used to counteract the effects of the atmosphere and PDI was used to reduce the stellar PSF and increase the contrast between the disc and the host star.

For HD~139614, GPI images were obtained from the Gemini Large Imaging with the Gemini Planet Imager Herbig/T Tauri Survey (Gemini-LIGHTS: \citealt{laws_2020, rich_2022}), data available from \citet{rich_2022_data}. The survey targeted stars with an infrared excess suggesting the presence of protoplanetary discs. Observations were carried out at the Gemini South observatory using coronagraphic spots with diameters of 184.7 and 246.7\,mas for the $J$ and $H$ band respectively. The pixel scale is 14.14\,mas/pixel. 

The data were processed using the Data Reduction Pipeline (DRP) \citep{maire_2010} with a python wrapper as explained in detail in \citet{rich_2022}. The resulting Stokes' cubes \citep{stokes_1851} contains total intensity ($I$), linear Stokes parameters ($Q$ and $U$), azimuthal Stokes parameters ($Q_{\phi}$ and $U_{\phi}$), and linearly-polarised intensity (LPI). This data cube can be flux calibrated from the ratio of the stellar magnitude and the flux from satellite spots to estimate the surface brightness of the disc (eg. \citealt{laws_2020}).

Further HD~139614 data were obtained from \citet{muro_arena_2020} using the SPEHRE instrument at the Very Large Telescope (VLT). Observations were carried out using the Infrared Dual-band Imager and Spectrograph (IRDIS: \citealt{dohlen_2008}) in differential polarimetric imaging mode \citep{langlois_2014}. A lyot coronagraph \citep{martinez_2009,carbillet_2011} with diameter 185\,mas was used to suppress the stellar light. The pixel scales are 12.263 and 12.251\,mas/pixel in the $J$ and $H$ band respectively \citep{marie_2016}. \citet{muro_arena_2020} followed the data reduction process described in \citet{boer_2016}. The released data consists of Stokes \textit{Q} and \textit{U} images which were transformed to $Q_{\phi}$ and $U_{\phi}$ (using equations in Appendix A of \citealt{monnier_2019}) which highlight the centrosymmetric nature of scattered polarised light.

\subsection{Analysis of Observations}
\label{sec: analysis_obs}

\begin{table*}
	\centering
	\caption{Table of the mean absolute surface brightness for the GPI data. Values are calculated for the outer disc (for an annulus of width $530-1000\,$mas) and for the bright ring (averaged between $20-90^\circ$ and $100-160\,$mas). The mean absolute surface brightness was calculated from the mean surface brightness of the annulus (from the $R^2$ corrected $Q_\phi$ image) scaled using the flux calibration scale factors from \citet{laws_2020}. The stellar magnitudes used to calculate the scale factors were taken from \citet{cutri_2003}.}
	\label{tab: surface_brightness}
	\begin{tabular}{lccccccc} 
		\hline
		Epoch & Band & Stellar Magnitude & \multicolumn{2}{c}{Mean Surface Brightness} & Scale Factor & \multicolumn{2}{c}{Mean Absolute Surface Brightness}\\
         & & & \multicolumn{2}{c}{[(ADU/coadd)/s]} & [mJy $\rm arcsec^{-2}/\rm(ADU/coadd)/s]$&\multicolumn{2}{c}{[mJy $\rm arcsec^{-2}$]}\\
         \hline
         &&&Outer disc& Bright-ring&&Outer disc& Bright-ring\\
		\hline
        2017 & J & $7.67\pm0.03$ & $761.2\pm8.6\phantom{0}$ & $48.3\pm2.1\phantom{0}$ & $2.8\pm0.6$ & $2131\pm457$&$135\pm30$\\
        2018 & H & $7.33\pm0.04$ & $941.3\pm10.9$ & $39.1\pm1.6\phantom{0}$& $3.1\pm0.5$ & $2918\pm472$& $121\pm20$\\
        2019 & J & $7.67\pm0.03$ & $573.3\pm7.0\phantom{0}$ &$36.3\pm1.7\phantom{0}$ & $1.8\pm0.4$ & $1032\pm230$& $\phantom{0}65\pm15$\\
		\hline
	\end{tabular}
\end{table*}

\begin{figure}
	\includegraphics[width=\columnwidth]{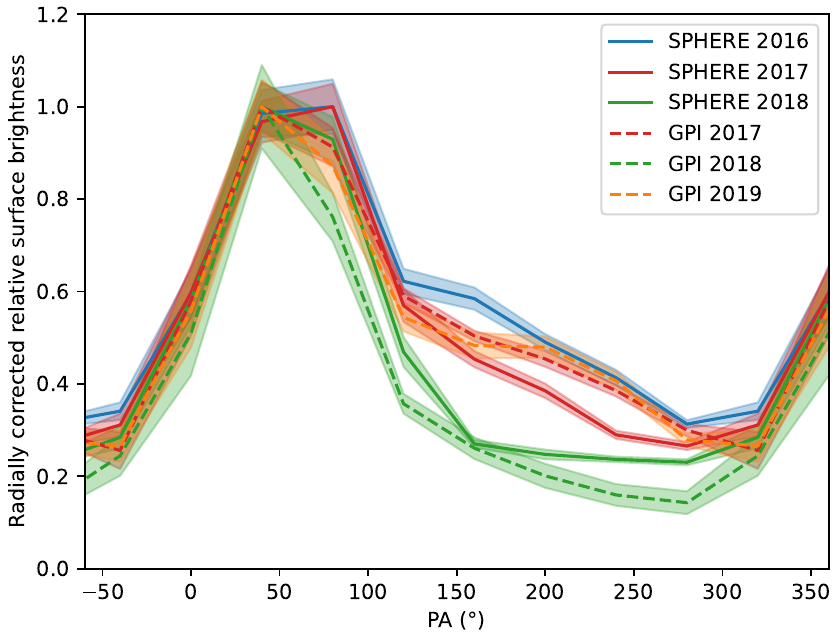}
    \caption{Azimuthal profiles of the bright ring for de-projected, $R^2$ scaled GPI (dashed line) and SPHERE (solid line) data. The 2016, 2017, 2018, and 2019 epochs are shown in blue, red, green, and orange respectively. The profiles are averaged between $13-21\,$au over 10 azimuthal bins. Each profile is individually normalised to the highest surface brightness value in that profile. Error bars are shown by the shaded regions.}
    \label{fig: azi_profile_bright_ring}
\end{figure}

\begin{figure}
	\includegraphics[width=\columnwidth]{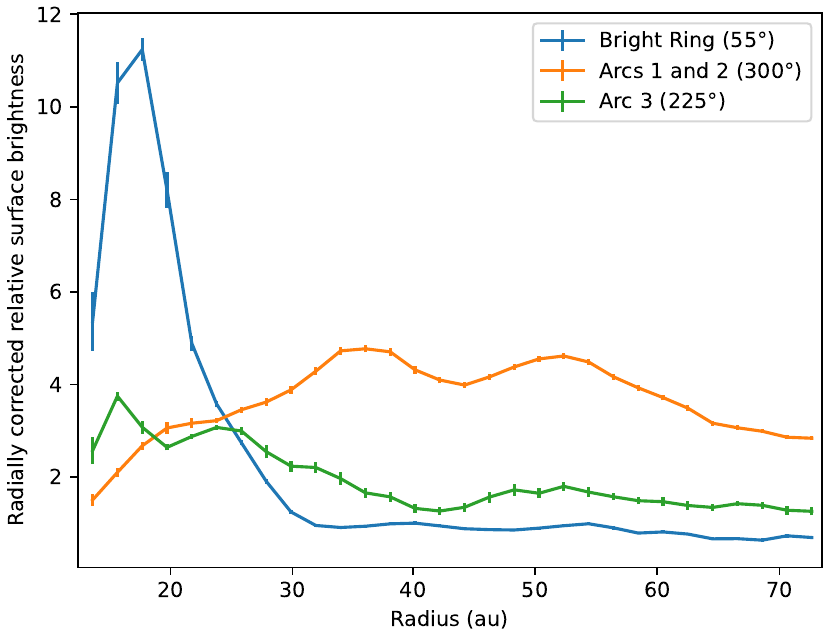}
    \caption{Radial profiles of the key features of the disc taken from the stacked SPHERE data (see Figure \ref{fig: stacked_sphere_qphi}). Three profiles are shown between $20-90^\circ$, $270-330^\circ$, and $210-240^\circ$ for the bright ring, arcs 1 and 2, and arc 3 respectively. The profiles were taken between $12\,$au (the edge of the coronagraph) and $73\,$au  each averaged over 30 radial bins. Error bars are shown. The legend shows which feature the profile shows and what angle it was centred on.}
    \label{fig: radial_profiles}
\end{figure}

\begin{figure*}
	\includegraphics[width=\textwidth]{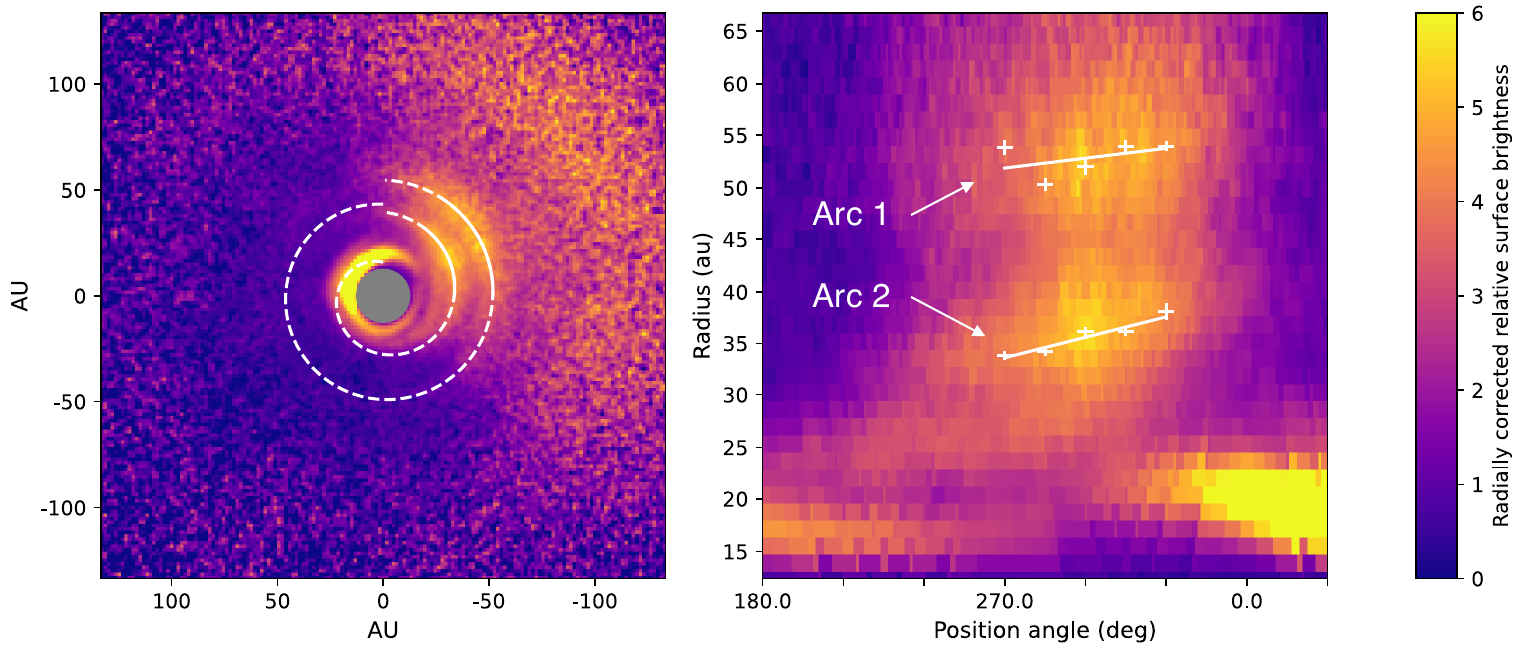}
    \caption{Left: De-projected, stacked $Q_{\phi}$ images of the SPHERE data (same as Figure \ref{fig: stacked_sphere_qphi}). The coronagraph is shown as a grey spot. Right: Polar plot of the image on the left (the bottom of the image starts at the edge of the coronagraph). The radial position of the peak surface brightness of the arcs are plotted for a certain angles (error bars are shown as a cross). The positions of the peak surface brightness are calculated from the averaged radial profiles. An Archimedean spiral (best fit line) is fit to the data and shown by a white line. The Archimedean spiral is also plotted on the Cartesian image (left) as a solid line between 270 and 330$^\circ$, as this is the range over which the arcs were analysed, and a dashed line continues the spirals beyond these angles. Each spiral is only plotted between 0-360$^\circ$.}
    \label{fig: spiral_fits}
\end{figure*}

\begin{table}
	\centering
	\caption{Table of  $\widetilde{\chi}^2$ and BIC values for the circular and spiral fits for arcs 1 (outer) and 2 (inner). The circular fit has one parameter, whereas the spiral fits both have two parameters. The values of the parameters calculated for the fits are listed in column 3.} 
	\label{tab:reduced_chi_squared}
	\begin{tabular}{lclcr} 
		\hline
		Arc & Model &Parameters&  $\widetilde{\chi}^2$&BIC\\
		\hline
        1 & Circular            & $a=53.0\pm 0.6\,\rm{au}$  & 6.8 & 26.6\\
        1 & Archimedean spiral  & $a=43.3\pm 9.9\,\rm{au}$  &6.9 & \phantom{0}5.1\\
        &&$b=0.03\pm 0.03\,\rm{au/^\circ}$&&\\
        1 & Logarithmic spiral  & $r_0=44.0\pm8.3\,\rm{au}$   &6.8 & \phantom{0}5.2 \\
        &&$k=0.035\pm 0.035$&&\\
        \hline
        2 & Circular            & $a=35.1\pm 0.6\,\rm{au}$   &9.1 & 38.0\\
        2 & Archimedean spiral  & $a=15.9\pm 3.8\,\rm{au}$   &1.3 & \phantom{0}7.1\\
        &&$b=0.07\pm 0.01\,\rm{au/^\circ}$&&\\
        2 & Logarithmic spiral  & $r_0=20.3\pm 2.1\,\rm{au}$   &1.3 & \phantom{0}7.1\\
        &&$k=0.107\pm 0.021$&&\\
		\hline
	\end{tabular}
\end{table}

The $Q_{\phi}$ images for HD~139614 are shown in Figure \ref{fig: deprojected_data_full_plot} with the SPHERE data on the left and the GPI data on the right (the $U_\phi$ images can be seen in Appendix \ref{sec: u_phi}). The images are de-projected so that the disc appears to be face-on, scaled with $R^2$, and normalised (see Section \ref{sec: normalisation}).

In order to see the structure of the disc more clearly, the images were de-projected to face-on using their inclination and position angles (PA). They were de-projected by creating a grid of coordinates for the de-projected image and transforming these coordinates to their positions in the original image. Pixels from the original image were then sampled for each coordinate, using cubic spline interpolation, to create the final de-projected image. 

The images were then scaled by $R^2$ to account for the intensity dropping off with the inverse square law, which improves detail at larger radii but accentuates noise (especially for the GPI images). This increase in noise at larger radii can most clearly be seen in the $U_{\rm \phi}$ images. The $U_{\rm \phi}$ images show very little signal suggesting that the scattered-light is dominated by the centrosymmetric pattern from the central object. In order to reduce noise a stacked image of the SPHERE epochs (as they are less noisy than the GPI data) was created in which the structures of the disc can be seen more clearly (Figure \ref{fig: stacked_sphere_qphi}). The images were first stacked, then de-projected, scaled with $R^2$, and then normalised.

\subsubsection{Normalisation}
\label{sec: normalisation}
After comparing the absolute surface brightness of the data we individually normalised all the scattered-light images (including the stacked image) for better comparison.

For the GPI data we calculated the absolute surface brightness of the disc in  mJy $\rm arcsec^{-2}$ as we have an absolute calibration of the flux. To compare the absolute surface brightness across epochs, we chose to calculate the mean surface brightness between $530-1000\,$mas ($\sim71-134\,$au) from the $Q_{\rm \phi}$ images (see Table \ref{tab: surface_brightness}). This radial range was chosen as it is outside the main features of the disc (arcs and bright ring), but close in to avoid noise (which is accentuated at large radii due to the $R^2$ scaling). The broad shadow is consistent across epochs so should not affect the comparison.

We additionally calculate the mean absolute surface brightness values for the brightest part of the ring between $20-90^\circ$ and $100-160\,$mas for comparison. This radial range was chosen as the averaged radial profile of the bright ring, from the stacked SPHERE data (see Figure \ref{fig: radial_profiles}), shows that surface brightness values above $60\%$ of the peak lie between $\sim100-160\,$mas. Similarly, the azimuthal range was chosen based on the averaged azimuthal profile of the stacked SPHERE image, where values above $80\%$ of the maximum surface brightness occur between $\sim20-90^\circ$, with angles measured anticlockwise from North ($0^\circ$). 

Scale factors to convert the mean surface brightness values to absolute surface brightness for each epoch were taken from \citet{laws_2020} (see Table \ref{tab: surface_brightness}). The scale factors were calculated using the ratio of the stellar magnitude and the flux from the satellite spots in the GPI data. The stellar magnitudes were from catalogues (eg. \citealt{cutri_2003}) rather than measured with the data, so do not take into account factors such as stellar variability. 

From Table \ref{tab: surface_brightness} we can see that the the average absolute surface brightness for the outer disc (between $530-1000\,$mas) is significantly lower for the 2019 epoch. We may expect a slight difference in absolute surface brightness between the $J$ and $H$ band images due to the difference in stellar magnitudes. We do not have absolute surface brightness measurements for the SPHERE data.

In order to better compare the data despite the differences in absolute surface brightness, the images were individually normalised to the mean surface brightness (between $530-1000\,$mas). The normalisation factor can be seen in the fourth column of Table \ref{tab: surface_brightness} for the GPI data. Since the images have been normalised, it is worth noting that the brightness in each image is relative to the mean surface brightness between $530-1000\,$mas. Therefore comparisons of brightness in this paper are relative to this region for each image.

\subsubsection{Surface brightness asymmetries} 

The surface brightness of the bright ring varies in two ways: azimuthally and temporally. The ring is brightest in azimuth between $\sim$20-90$^{\circ}$ and dimmest in the West. At its dimmest, the ring is about a factor of 5 dimmer than at its brightest, as measured from an averaged azimuthal profile of the bright ring (between $13-21\,$au) from the stacked image.

The temporal variation in the bright ring can be seen from the azimuthal profiles of the bright ring for each epoch. Figure \ref{fig: azi_profile_bright_ring} shows profiles for each epoch (coloured lines) with the SPHERE data shown by dotted lines and the GPI data by solid lines. Error bars are shown by shaded regions. The profiles were created by taking 10 azimuthal bins, spaced evenly between 0-360$^{\circ}$. Pixels were taken between $13-21\,$au for each bin and averaged to find the average surface brightness at that angle. Errors were calculated using the standard error of the mean and each profile was individually normalised to the maximum surface brightness. 

The position of the peak of the bright ring is consistent across epochs. However, it is clear from the profiles that the brightness distribution shifts for the 2018 epoch: between $\sim$125-300$^\circ$ the bright ring is dimmer for the 2018 epoch than the other epochs. This can also been seen in the $Q_{\phi}$ images for both the SPHERE and GPI data in Figure \ref{fig: deprojected_data_full_plot}; the bright ring is brighter (compared to the brightest part of the ring) between $\sim$125-300$^\circ$ for the 2016 and 2017 epochs, then dims for the 2018 epochs, returning to its original brightness (compared to the brightest part) for the 2019 epoch.

An azimuthal asymmetry in brightness can also be seen in the outer disc which has a broad shadow between $\sim$0-230$^{\circ}$ (see Figure \ref{fig: stacked_sphere_qphi}). The outer disc can be seen in the scattered-light images out to a radius of $220\,$au and the shadow can be seen across $\sim2/3$ of the disc. The position of the shadow is consistent across all epochs, but it does not line up with the position of the shadow cast across the bright ring. 

From Table \ref{tab: surface_brightness} it is clear that the stellar magnitude is lowest for the 2018 epoch (so the star is brightest), therefore it may be suggested that bright ring is brightest for this epoch and so the shadow simply appears to be dimmer as we are comparing to a brighter ring. However from the mean absolute surface brightness for the brightest part of the bright ring (rightmost column) it is clear that the ring is not brightest for the 2018 epoch (unlike the outer disc). Therefore it is likely that the brightness distribution of the bright ring does change for the 2018 epoch.

In the bright part of the disc (between $\sim$230-360$^{\circ}$) two bright arc-like features are visible at $\sim44-57\,$au (labelled arc 1) and $\sim31-44\,$au (labelled arc 2). There is also a third, fainter, arc (in the South-West) at $\sim21-29\,$au which can be seen most clearly in the stacked SPHERE data.

The radial profiles for the arcs and the bright ring are shown in Figure \ref{fig: radial_profiles}. The profiles were created from the stacked SPHERE image between $12-73\,$au. Different angle ranges were chosen to show specific features of the disc. Each wedge was split into 30 evenly spaced radial bins over which the surface brightness was averaged. The centre of each bin was taken to be the radius for that surface brightness value. The error bars were calculated over all pixels within the bin from the standard error of the mean surface brightness.

From the radial profile centred on $55^\circ$ (blue line) there is large spike in surface brightness centred on $\sim17\,$au due to the inner bright ring which can be seen in the scattered-light images. The profile centred on $300^\circ$ (orange line) has two peaks at $\sim52\,$au and $\sim35\,$au due to two arcs in the North-West of the scattered-light images. Although the arcs are still bright, the peak surface brightness of the bright ring is a factor of $\sim2.4$ times brighter than the peak surface brightness of the arcs. Finally, the profile centred on $225^\circ$ (green line) shows a peak at $\sim16\,$au due to the dimmer side of the bright ring, but also a peak at $\sim24\,$au from the third fainter arc.

\subsubsection{Arcs}
The two bright arcs were analysed to determine if they are likely to be spiral or circular in nature using the stacked $Q_{\phi}$ image. The third arc (labelled in Figure \ref{fig: stacked_sphere_qphi}) was not considered due to poor signal-to-noise and the data were too noisy to determine the shapes of the arcs for each individual epoch.

To determine the nature of the arcs, the stacked $Q_{\phi}$ image was plotted in polar coordinates, as shown in Figure \ref{fig: spiral_fits} which shows the stacked, de-projected image of the SPHERE data in Cartesian coordinates (left) and polar coordinates (right). To create the polar plot, a list of $r$ and $\theta$ values, spaced evenly (every 1\,mas and $1^\circ$), was created. These polar coordinates were transformed to find the Cartesian coordinates and the surface brightness values were sampled from the Cartesian image.

Radial profiles were taken at various angles, and from these, the position of the peak surface brightness for each arc was calculated for each angle. The radial profiles were produced by taking five azimuthal cuts, with a width of $6^{\circ}$, at regularly spaced angles between $270-330^\circ$. These were then averaged over 30 radial bins (between $12-73\,$au) to get five averaged radial profiles. The radial position of the peak surface brightness was calculated from a slice of the radial profile, taken from the minimum between the two arcs and the same width the other side of the arc which was being analysed. A slice was taken so that the radial position of the peak surface brightness was not skewed. 

The position of the peak surface brightness was taken from the radial profile slice using equation:
\begin{align}
    \bar d=\frac{\sum df}{\sum f}
    \label{eq: moment_analysis}
\end{align}
where $\bar d$ is the radial position of the peak surface brightness, $d$ is the radial position of each data point, and $f$ is the corresponding surface brightness value of that data point. 

The values of $\bar d$ were then plotted for each angle on the polar plot of the stacked SPHERE data (right of Figure \ref{fig: spiral_fits}) with the error bars shown as crosses. Vertical error bars were calculated from the errors (in $f$ and $d$) from the radial profile and propagated through Equation \ref{eq: moment_analysis}. Horizontal error bars were calculated from the width of the wedge used to create the radial cuts. 

A best-fit line was calculated for the arcs from the position of the peak surface brightness plotted on the polar plot. From this we can determine the nature of the arcs: if the line fit to an arc has zero gradient on a polar plot then the arc is circular in nature, but if it is sloped the arc is spiral. A straight best-fit line was fitted to each set of points using non-linear least squares to find the optimal function by minimising $\chi^2$. This method included the y-axis error bars to fit the function. 

This best-fit line for the peak brightness of the arcs (see Figure \ref{fig: spiral_fits}) is an Archimedean spiral as this is a straight line in a graph of $r$ vs $\theta$: the equation for this spiral is $r=a+b\theta$, where $a$ and $b$ are constants (see Table \ref{tab:reduced_chi_squared} for values of these parameters). The spiral (or best fit line) was also plotted on the Cartesian plot for the stacked SPHERE data (left of Figure \ref{fig: spiral_fits}). The solid lines show the spiral fit between 270-330$^\circ$, where the analysis from the radial peaks was carried out, and the dashed lines show a continuation of the spiral fits.

A logarithmic spiral was also fitted to the arcs to estimate their pitch angle as this spiral has a constant pitch angle. The equation for a logarithmic spiral is $r=r_0e^{k\theta}$ where $r_0$ is a constant, $\theta$ is in radians, and $k$ is related to the pitch angle, $\phi$, of the spiral by $\phi=\arctan{k}$ (see Table \ref{tab:reduced_chi_squared} for values of $r_0$ and $k$). This equation was fitted to the points of the peak surface brightness for each arc using the same technique as for the best-fit line (including the y-axis errors) to get the parameters $r_0$ and $k$. From $k$, the pitch angles were estimated to be $2.0\pm2.0^\circ$ and $6.1\pm1.2^\circ$ for arcs 1 and 2 respectively. 

Finally, a circle was fitted to the arcs (a line with zero gradient on the polar plot) in order to determine if the spiral fits were an improvement on a circular fit. The value of reduced $\chi^2$ was calculated to analyse the goodness of fit of the spiral fits using equation:
\begin{align}
    \widetilde{\chi}^2 = \frac{\chi^2}{n}
	\label{eq: reduced_chi_squared}
\end{align}
where $n$ is the number of degrees of freedom of the system, taken to be the number of data points minus the number of parameters of the model. 

For the spiral fits for the arcs the $\widetilde{\chi}^2$ can be seen in Table \ref{tab:reduced_chi_squared}. For arc 1 (the outer arc) we can see that the $\widetilde{\chi}^2 \sim7$ for both spiral fits and the circular fit. We would expect the $\widetilde{\chi}^2$ value to be less for the spiral fits than the circular fit as these models have an additional parameter. The fact that the values are the same suggests that the spiral fits are not an improvement on the circular fit as the extra degree of freedom did not improve the value of $\widetilde{\chi}^2$ significantly. Since $\widetilde{\chi}^2 \gg1$ the models are statistically rejected at the $1\%$ significance levels.

However, we show that arc 2 is likely to be spiral in nature as $\widetilde{\chi}^2 \sim1$ for the spiral fits suggesting that they are a good fit. It is unsurprising that the values for the spiral models are very similar as the fits are almost indistinguishable for such as small angle range. The spiral fits also show a significant improvement over the circular fit for which $\widetilde{\chi}^2 \sim9$ so it is statistically rejected at the $1\%$ significance level.

To test if the spiral model for arc 2 is appropriate we need to see if there is a significant improvement in $\widetilde{\chi}^2$ when the number of parameters of the model is increased. To do this we use the Bayesian information criterion \citep{schwarz_1978}, BIC, which is weighted by the number of model parameters, given by:
\begin{align}
    \mathrm{BIC}=\chi^2+k\ln{n}
\end{align}
where $k$ is the number of model parameters, and $n$ is the number of data points. Lower $\mathrm{BIC}$ values indicate a better balance between the model fit and its number of parameters. For arc 2, $\mathrm{BIC}=38.0$ for the circular fit and $7.1$ for the spiral fits. Therefore $\Delta\mathrm{BIC}=30.9$ so the spiral fit is a significant improvement on the circular fit (despite its extra model parameter).

We note that for arc 1 the data point at $270^\circ$ appears anomalous and does not fit the slope of the other data points. Removing this data point we measure the pitch angle of the spiral to be $5.0\pm1.2^\circ$ which is close to the pitch angle measured for arc 2. The $\widetilde{\chi}^2$ values are $1.1$, $1.2$, and $8.4$ for the Archimedean, logarithmic, and circular fits respectively. The circular fit is statistically rejected at the $1\%$ significance level, but the spiral fits (with $\widetilde{\chi}^2\sim1$) are good fits. The BIC values are $5.1$, $5.2$, and $26.6$ for the Archimedean, logarithmic, and circular fits respectively. Therefore $\Delta \rm BIC \sim20$ between the spiral and circular fits which is strong evidence that the spiral fits are an improvement on the circular fit. However, we note that we do not have an objective reason to reject this data point.

\subsection{Discussion of Observations}
\label{sec: discussion_obs}

\subsubsection{Temporally variable bright ring}
\label{sec: discussion_obs_bright_ring}
We have shown that the azimuthal brightness distribution of the inner ring varies with time. From Figures \ref{fig: deprojected_data_full_plot}, and \ref{fig: azi_profile_bright_ring} we can see that for the 2018 epoch the brightness distribution of the rings has shifted and the ring is dimmer (compared to its brightest) between $\sim$125-300$^\circ$. \citet{muro_arena_2020} observed this variation in the SPHERE data, but they interpreted it as ``variations in the centering of the images due to the stellar jitter behind the coronagraph'' causing the bright ring to be off-centre. However, since we see this variation in two independent data sets taken with different instruments we suggest that it is a real, physical feature and it is unlikely to be due to centering issues. 

It is unlikely that this temporal variation in the brightness of the inner ring is simply due to a clump of dust in the ring orbiting at the inner ring radius (eg. \citealt{rich_2019}). At the radius of the bright ring, $\sim$17 au, on a Keplerian orbit, the time period of one orbit would be $\sim58\,$years and so the bright ring would not vary significantly over the three years which we have observations. 

The 2016 and 2017 epochs show the brighter inner ring, then the ring dims in the South (relative to its brightest) for the 2018 epoch, then is bright again for the 2019 epoch. There are $\sim13$ months between the later 2017 epoch (where there is no variation in the shadow) and earlier 2018 epoch (see Table \ref{tab:observations}) and $\sim11$ months between the later 2018 epoch and the 2019 epoch. This shows that the variation in brightness could have a timescale of less than $11\,$months. Something on a Keplerian orbit with a timescale of 11 months would have an orbital radius of $\sim$1\,au ($8\,$mas). This strongly suggests that the shadow could be cast by a structure close to the host star. 

Interestingly, the 2016 epoch and earlier 2017 epoch are about a year apart, and the brightness distribution does not appear to vary across these epochs. This suggests that the shadow could be a one off event or that it is varying on a timescale shorter than a year. We know that the variation lasts at least $\sim$2 weeks as the 2018 GPI and SPHERE observations are 13 days apart, however we cannot constrain the minimum timescale of the variation further. 

Further scattered-light observations at multiple epochs could be used to constrain the timescale of the variability of the bright ring. Scattered-light observations at the same epochs as interferometric observations, would allow exploration of how variations in the inner disc structure influence shadowing patterns in the outer disc. Additionally disc kinematics obtained from molecular line observations could be used to analyse the structure of the inner disc.

\subsubsection{Spiral arcs}
\label{sec: discussion_spiral_arcs}

We have shown that the two arcs may be spiral shaped (as previously suggested by \citealt{laws_2020}). The Archimedean spiral fit to the arcs can be seen in the left of Figure \ref{fig: spiral_fits} for the stacked SPHERE data. For arc 2, the Archimedean fit follows the shape of the arc well and seems to trace the shape of the potential third inner arc too. Interestingly, the continuation of the spiral also passes through the bright part of the inner ring.

We estimated pitch angle for arc 2 to be $6.1\pm1.2^\circ$ suggesting that the spiral is fairly tightly wound. Spirals features such as this could be formed from a variety of mechanisms, such as infall onto the disc from the star forming environment \citep{lesur_2015, kuznetsova_2022}, which creates fairly tightly wound spirals with a pitch angle of $3-9^\circ$ when $H/R = 0.05-0.15$ \citep{hennebelle_2016, bae2023structured}. 

The spiral could also be caused by buoyancy waves created by a perturbing companion (eg. \citealt{zhu_2012}). These spirals have a pitch angle of $\lesssim10$ close to the perturbing companion \citep{bae_2021}, therefore spiral nature of arc 2 could signal the presence of a companion. It is possible that such a companion could also explain the shadowing if it could be causing a warp the inner disc (eg. for HD142527, \citet{price_2018} showed that shadows and spirals could be explained by the presence of a binary companion).

A common explanation for spirals is gravitational instability, however this disc is older and less massive than young protoplanetary discs, and so the spirals are unlikely to be due to gravitational instability as the disc-to-star mass ratio ($M_D/M_*$) is too small \citep{toomre_1964,forgan_2011,hall_2019}. 

Spiral waves may also be triggered by shadows cast across the disc \citep{montesinos_2016}. The difference in temperature between the shadowed and illuminated parts of the disc can create pressure gradients which induce spiral waves. These spiral waves have fairly low pitch angles between $\sim11-22^\circ$. Spiral waves can also be induced by precessing shadows cast by an inner warp  \citep{montesinos_2018}. However, hydrodynamical simulations of spiral waves induced by shadows focus on two narrow shadows, due to an inclined disc with a large inclination, rather than the broad shadowing seen for HD~139614.\\\\ The consistent and temporally-invariant feature of the disc is the broad shadow. The presence of the bright ring, with an azimuthal brightness variation that is not apparently aligned with broad shadow is an additional feature that warrants explanation. In the following section we attempt to construct a simple, physical model that explains both these features.

\section{Modelling}
\label{sec: modelling}

Previous modelling of HD~139614 by \citet{matter_2016} explored the structure of the inner disc. Using a combination of interferometric observations and the SED to constrain their model, they confirmed the presence of a gap in the dust between $\sim2.5-5.7$au. 

\citet{muro_arena_2020} extended this model to try and explain large scale features, such as the broad shadow (across $\sim230^\circ$) seen in the outer disc, using a series of misaligned annuli. Two misaligned annuli were used to produce a sufficiently broad shadow since a single annulus could not result in a  shadow broader than $180^\circ$. Additionally, to achieve the broad shadowing seen in the outer disc without producing two opposing shadows (see Section \ref{sec: introduction}), the rings had small misalignments relative to each other and the outer disc.  To reproduce the arcs, the scale height in the outermost ring was iteratively adjusted to best fit the surface brightness radial profile of the illuminated part of the disc. The model accurately reproduced the arcs, the broad shadow across the outer disc, and the bright inner ring in the scattered-light images. Also, the SED from their model matched the observed SED.

Despite the success of the Muro-Arena model in explaining the observations, it is geometric rather than physical. The misaligned annuli would exchange angular momentum via internal torques and would naturally evolve towards a co-planar state in the absence of external torques.

Hence the fact that we infer that there is a warp or misalignment in the disc of HD~139614 suggests that there is an external torque driving this warp/misalignment such as a planet orbiting out of the plane of the disc \citep{nealon_2018}. There is already indirect evidence for a planet in the disc (see Section \ref{sec: intro_hd139614}), and below we explore the possibility that this putative planet is the source of an external torque that shapes the inner disc and causes the misalignments (eg. \citealt{nealon_2018, nealon_2019, young_2023}) that produce the broad shadow and bright ring.

We use 1D warp equations to evolve our disc model with time (Section \ref{sec: 1d_disc_model}) and combine this  with a fast radiative transfer code to produce scattered-light observations of our model (Section \ref{sec: radiative_transfer}). We setup our disc as a series of three annuli separated by gaps. We then perturb the inner two annuli by a planet on an inclined orbit which causes the annuli to tilt and precess, casting shadows on the outer disc (Section \ref{sec: model setup}). 

Our aim is to create a physically motivated model which produces a shadow across more than half of the disc with a bright ring in the inner disc. We emphasise that this is not an attempt to reproduce the intricacies of the disc in detail.

\subsection{1D warp model}
\label{sec: 1d_disc_model}

In order to model our disc we use a 1D approach as this is computationally much less expensive than a full hydrodynamic treatment. We treat the disc as a series of symmetric annuli defined by a radial grid. We can then evolve the internal torque of the disc, $\textbf{\textit{G}}$ and the angular momentum vector of the disc, \textbf{\textit{l}} with time, $t$, using equations 12 and 13 from \citet{lubow_oglivie_2000} (or 4 and 5 from \citealt{martin_2019}) which are applicable for small amplitude warps in the wave-like regime:
\begin{align}
    \frac{\partial \textbf{\textit{G}}}{\partial t} - \omega\textbf{\textit{l}} \times\textbf{\textit{G}} + \alpha\Omega\textbf{\textit{G}} = \frac{\Sigma H^2R^3\Omega^3}{4}\frac{\partial \textbf{\textit{l}}}{\partial R}
	\label{eq: internal_torque}
\end{align}
and 
\begin{align}
    \Sigma R^2\Omega\frac{\partial \textbf{\textit{l}}}{\partial t} = \frac{1}{R}\frac{\partial \textbf{\textit{G}}}{\partial R} + \textbf{\textit{T}}
	\label{eq: angular_momentum}
\end{align}
where $\Omega$ is the angular frequency of the disc at radius $R$, $\Sigma$ is the surface density of the disc at radius $R$, $H$ is the height of the disc from the midplane, $\textbf{\textit{T}}$ is an external torque  (eg. due to a perturbing planet), and $\omega$ is the apsidal precession frequency in the plane of the disc given by:
\begin{align}
    \omega = \frac{\Omega^2-\kappa^2}{2\Omega}
	\label{eq: apsidal_precession_frequency}
\end{align}
where $\kappa$ is the epicyclic frequency.

If we assume that the mass of a perturbing planet is small relative to the stellar mass, the main gravitational force on the disc is from star and any perturbations from Keplerian motion due to the planet are negligible. We can therefore assume that the disc is moving with Keplerian frequency ($\Omega(R)=\sqrt{GM/R^3}$) and so $\Omega=\kappa$ and $\omega=0$ and we can ignore the second term of Equation \ref{eq: internal_torque}. This is often assumed for the wave-like regime \citep{papaloizou_lin_1995} which is applicable to protoplanetary discs. Note that if there is sufficient deviation from Keplerian motion, such as a warp induced by a massive binary \citet{lubow_oglivie_2000} (although only the inner edge of the disc is affected) or a warp induced by a rotating black hole \citep{Lubow_2002}, then the apsidal precession term is important, but it is not relevant for the mass of the perturber in this paper (see Section \ref{sec: model setup}). 

We model the external torque on the disc due to a perturbing planet using equation 22 from \citet{lubow_oglivie_2000}:
\begin{align}
     \textbf{\textit{T}} = \frac{GM_{\rm p}\Sigma R}{2R^4_{\rm p}}\left[b^{(1)}_{3/2}\left(\frac{R}{R_{\rm p}}\right)\right](\textbf{\textit{R}}_{\rm p}\cdot\textbf{\textit{l}})(\textbf{\textit{R}}_{\rm p}\times\textbf{\textit{l}})
	\label{eq: lubog_torque}
\end{align}
where $G$ is the gravitational constant, $M_{\rm p}$ is the mass of the perturber, $\textbf{\textit{R}}_{\rm p}$ is the position vector of the perturber at time $t$, and $b$ is the Laplace coefficient given by:
\begin{align}
     b_\gamma^{(m)}(x)= \frac{2}{\pi}\int^\pi_0 \cos(m\phi)(1+x^2-2x\cos\phi)^{-\gamma}d\phi\ .
	\label{eq: laplace}
\end{align}
This  allows us to calculate the external torque due to the planet at each annulus of the disc simply from the position of the planet ($\textbf{\textit{R}}_{\rm p}$) at time $t$. 

We set up the disc to be in $xy$-plane on a staggered grid with $\textbf{\textit{G}}$ defined at half grid points at the edge of each annulus (as it tells us how each annulus affects the adjacent ones), and $\textbf{\textit{l}}$ defined at full grid points at the centre of each annulus. We can calculate the tilt (at each annulus) from $\textbf{\textit{l}}$ (see Section \ref{sec:precession}).

We  solve Equations \ref{eq: internal_torque} and \ref{eq: angular_momentum} using the numerical method outlined in section 4.4 of \citet{Lubow_2002} (and explained in detail in \citealt{milsom_2026}). We set up $\textbf{\textit{l}}$ and $\textbf{\textit{G}}$ at full and intermediate half time steps. From Equation \ref{eq: internal_torque}, a full time step for $\textbf{\textit{G}}$ is calculated from: the previous full time step of $\textbf{\textit{G}}$ for the viscous term (Euler method), and from the intermediate time step of $\textbf{\textit{l}}$ for the radial derivative (using numerical differencing). A full time step for $\textbf{\textit{l}}$ is calculated using Equation \ref{eq: angular_momentum} from: an intermediate time step of $\textbf{\textit{G}}$ for the radial derivative (using numerical differencing), and an intermediate time step of $\textbf{\textit{l}}$ for the external torque term (leapfrog method). Intermediate time steps can then be calculated from full time steps (and the previous intermediate time step) and so on.

\subsection{Model setup}
\label{sec: model setup}
\begin{figure}
	\includegraphics[width=\columnwidth]{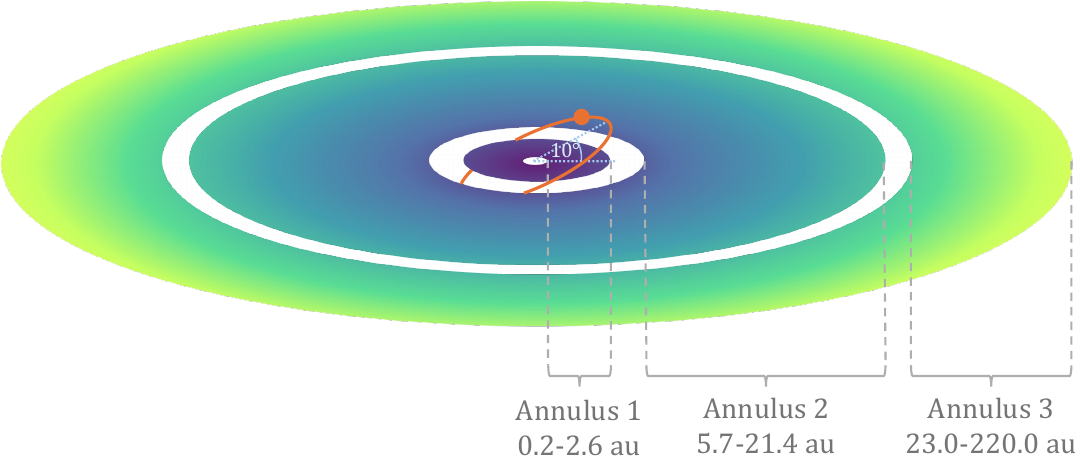}
    \caption{Diagram showing the setup of our model for the inner disc. The disc is made up of three annuli separated by gaps: annulus 1 ($0.2-2.6\,$au), annulus 2 ($5.7-21.4\,$au), and annulus 3 ($23.0-220\,$au) which is not fully shown. Between annuli 1 and 2 we place a $3M_{\rm jup}$ planet inclined by $10^\circ$ (orbit shown in orange). The planet perturbs annuli 1 and 2 which independently tilt and precess round. Annulus 3, which is further from the planet, stays relatively flat. We note that this diagram is not to scale and only the inner part of annulus 3 is shown.}
    \label{fig: model_setup}
\end{figure}

We set up our model using parameters consistent with observations and previous modelling of HD~139614. We take the mass of the central star as $1.481M_\odot$ (see Table \ref{tab:stellar_params}), the dust mass of the disc as $41.7M_\oplus=1.252\times10^{-4}M_\odot$ (see Table \ref{tab:disc_params}), and assume a dust to gas ratio of 1:100. The surface density profile for our model has a tapered inner edge given by \citep{pringle_1981, nealon_2018}:
\begin{align}
    \Sigma(R)= \Sigma_0 R^{-1}\left(1-\sqrt{\frac{R_{\rm in}}{R}} \right)
	\label{eq: surface_density}
\end{align}
where $\Sigma_0$ is a scaling parameter to be determined by the disc mass, and $R_{\rm in}$ is the inner radius of the disc. 

We adopt a flared disc structure in which the scale height ($H/R$) is given by the power law:
\begin{align}
    H = H_0\left( \frac{R}{R_0} \right)^\beta
	\label{eq: hr_power_law}
\end{align}
where $H_0$ is the height of the disc from the midplane at $R=R_0=100\,$au, and $\beta$ is a flaring index. We set $H_0=10\,$au and $\beta=1.025$. Note that this is consistent the low flaring found by \citet{muro_arena_2020} in order to have a similar surface brightness profile. Further motivation for a flatter disc is that it increases the surface brightness of the bright ring relative to the outer disc (see Section \ref{sec: model_results}). We note that different $H/R$ values could reproduce similar results, but it is beyond the scope of this work to fit for this parameter (see Section \ref{sec: model_caveats}).

For protoplanetary discs we can assume that we are in the wavelike regime ($\alpha<H/R$) so we adopt a low viscosity of $\alpha=0.001$ and that the disc has Keplerian rotation so $\Omega=\sqrt{GM_*/R^3}$. The radial grid to extends from $0.2$ \citep{matter_2014} to $220\,$au (as this is the extend of the disc in scattered-light) on a logarithmic scale with 1000 radial grid points. 

To produce a broad shadow across more than half of the disc we model the disc as three annuli (separated by gaps) which can tilt and precess independently. We define the radial extent of each annulus based on observations of HD~139614.  \citet{matter_2016} suggest there is a gap in the very inner disc between $2.6-5.7\,$au (see Table \ref{tab:disc_params}) which we adopt as the boundary separating annulus 1 and 2. 

We chose the outer gap (separating annulus 2 and 3) to be between $21.4-23\,$au so that annulus 2 ends at the outer edge of the bright ring. We set $23\,$au as the outer radius of the gap as it is large enough for annuli 2 and 3 to be separate, but small enough that the gap is not obvious in scattered-light (as we do not see a gap at this radius in the observations - see model caveats in Section \ref{sec: model_caveats}). Therefore annulus 1 extends from $0.2-2.6\,$au, annulus 2 from $5.7-21.4\,$au, and annulus 3 (the outer disc) from $23.0-220.0\,$au (see Figure \ref{fig: model_setup}).

We place a $3M_{\rm jup}$ planet at $4\,$au (in the gap between annulus 1 and 2) as the presence of such a planet was suggested by \citet{matter_2016, carmona_2017}, on an orbit inclined by $10^\circ$. We assume a circular orbit, and tilt the planet about the $y$-axis. The planet perturbs the inner annuli so they tilt and precess while the largest, outer annulus, which is further from the planet, stays relatively flat. We enforce boundary conditions at the edge of each annulus such that $\mathrm{d}\textbf{\textit{l}}/\mathrm{d}r=0$ and $\textbf{\textit{G}}=0$.

We tested other inclinations for the planet orbit ($i_{\rm p}=2^\circ,5^\circ,10^\circ,20^\circ,30^\circ,40^\circ$), but found that for $i_{\rm p}>10^\circ$ the annuli tilted enough to produce two narrow shadows opposite each other rather than broad shadows seen at lower inclinations. For $i_{\rm p}=5^\circ$ we see similar results to $i_{\rm p}=10^\circ$ (see Appendix \ref{sec: beta_5}), but the shadows are more prominent for $i_{\rm p}=10^\circ$ so we focus on these results for clarity. Models with $i_{\rm p}=2^\circ$ produced only very faint shadows.

To prevent a warp wave due to suddenly introducing a torque due to a planet we initially set the planet mass to zero and increase it linearly with time up to $t_0$ according to the equation:
\begin{align}
    M_{\rm p}=\frac{t}{t_0}M_{\rm f}
	\label{eq: mass_increase}
\end{align}
where $M_{\rm p}$ is the mass of the planet at time $t$ and $M_{\rm f}$ is the final mass of the planet ($3M_{\rm jup}$). We set $t_0$ to $1500\,$ rotations of the planet ($9900\,$years).

To ensure that our numerical method is stable we us the Courant-Friedrichs–Lewy condition \citep{courant_1928} to determine our time-step:
\begin{align}
    C = u\frac{\Delta t}{\Delta R}
	\label{eq: courant_condition}
\end{align}
where $C$ is the Courant number, $\Delta t$ is the time-step, $\Delta R$ is the distance between radial grid points, and $u$ is the magnitude of velocity of either the propagating perturbation or perturbing planet (whichever is higher). For the numerical method to be stable, $C\leq1$ so we calculate $\Delta t$ for $C=1$ and divide by 3 to ensure that our time-step is low enough.

\subsection{Radiative transfer code}
\label{sec: radiative_transfer}
In order to compare our model to scattered-light observations, we must model the interaction of stellar light with the disc to produce scattered-light images from our model. We produce images for many time steps of our model (which we run for tens of thousands of years) to visualise how the shadows evolve with time and see if any time steps have features similar to observations of HD~139614. However, existing radiative transfer codes that use Monte Carlo methods take minutes to run per image (e.g. \citealt{Harries_2019, Sheehan_2017,williams_2025}). Therefore we use a fast radiative transfer code explained in detail in \citet{milsom_2026}. The code takes seconds to run on a laptop making it approximately two orders of magnitude faster than a Monte Carlo radiative transfer code.

The code is a monochromatic radiative transfer code that models scattered-light using only a single-scattering approximation. It does not include a radiative equilibrium calculation, or thermal emission. However, for scattered-light images taken using adaptive optics and a coronagraph (to block light from the host star) direct thermal emission from the inner disc is obscured and we show that the fast radiative transfer code is a good approximation to a full radiative transfer calculation in \citet{milsom_2026}.

\subsection{Model analysis and results}
\label{sec: model_results}
\begin{figure*}	\includegraphics[width=0.88\textwidth]{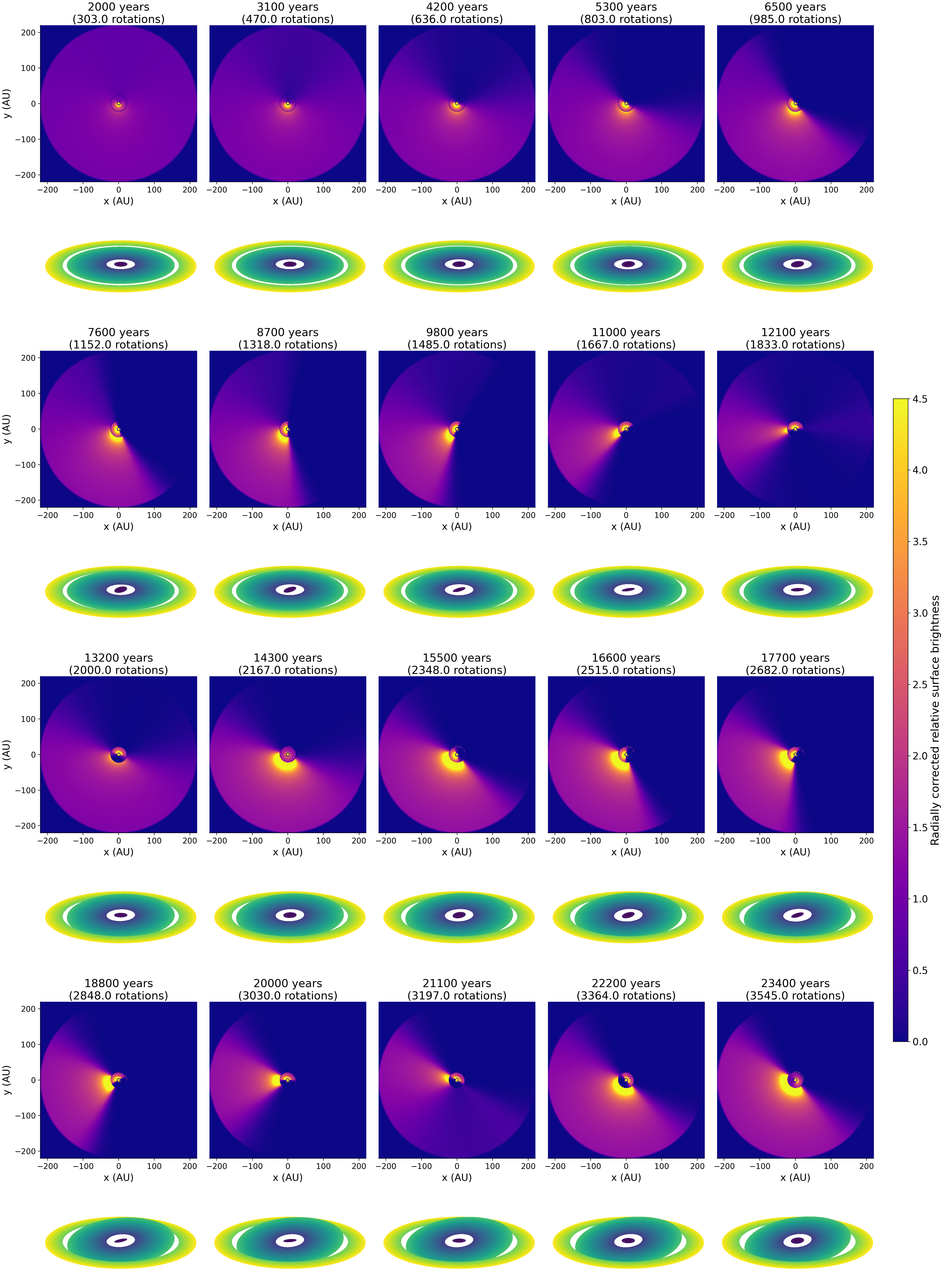}
    \caption{Snapshots of shadows in scattered-light images from our model of two precessing rings perturbed by an inclined planet. Rows show scattered-light images at different times (columns) with the orientation of the inner annuli shown by the diagram below each image. In the diagrams the disc has been inclined by $70^\circ$ (where $0^\circ$ is face-on) and only part of the outer annulus (annulus 3) is shown. The scattered-light images have been scaled with $R^2$ and normalised the the highest normalisation factor of the images. Broad shadows are cast across the outer disc (annulus 3) by annulus 1 and 2. Annulus 2 appears as a bright ring and it is only shadowed by annulus 1. Annulus 2 precesses slowly so the shadow it casts is in the North-West for every image, where as the shadow cast by annuls 1 precesses round. Images are shown for two precession periods of annulus 1.}
    \label{fig: model_results}
\end{figure*}

\begin{figure*}
	\includegraphics[width=\textwidth]{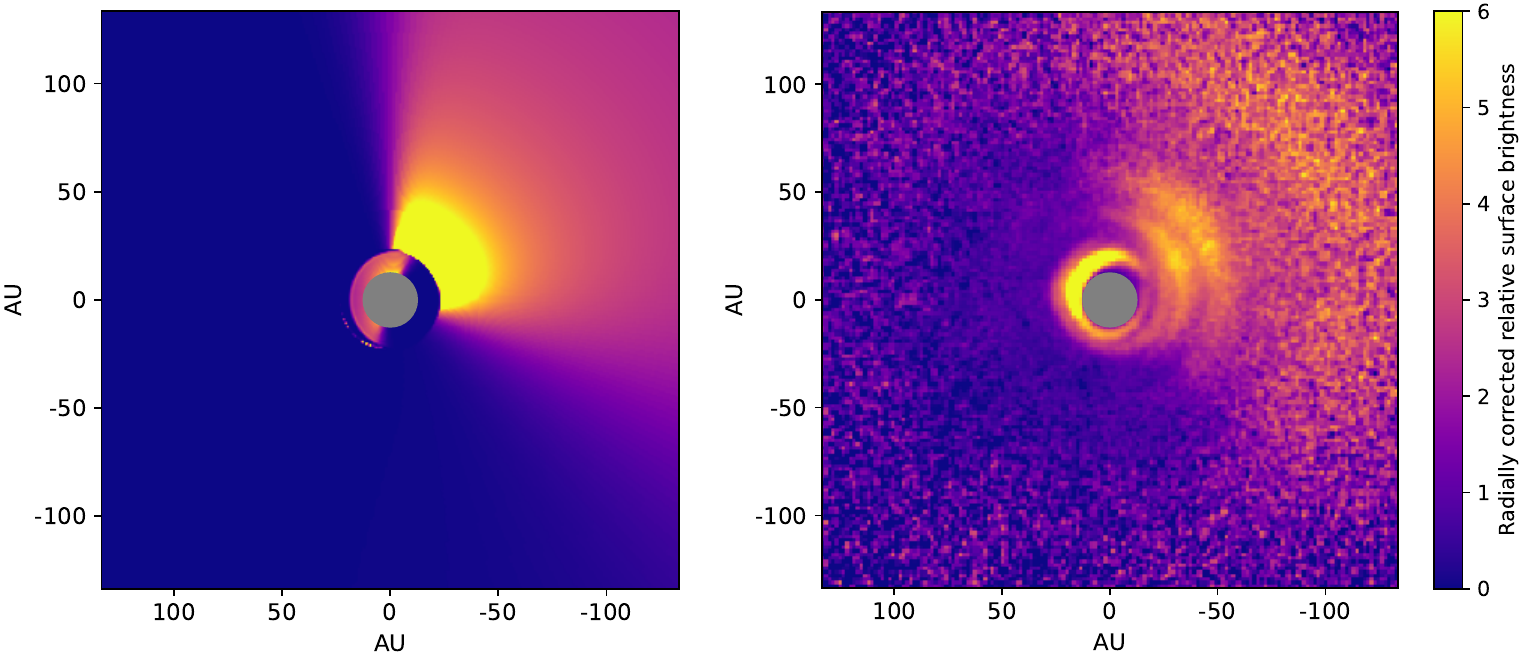}
    \caption{Snapshot of our model (left) with the HD~139614 observations (right). The observations are the de-projected, stacked $Q_{\phi}$ image of the SPHERE data which has been $R^2$ scaled and normalised (image from Figure \ref{fig: stacked_sphere_qphi}). The scattered-light image from the model was taken at 7242 rotations of the planet ($47,800\,$years), $R^2$ scaled and normalised. The image has been zoomed in to match the scaling of the observations and rotated clockwise $165^\circ$ so that the position of the shadow matches. A coronagraph has been added (grey) to match the observations. We can clearly see a broad shadow across two thirds of the disc and an inner bright ring which is not aligned with the bright region in the outer disc. The position of the shadow and bright ring match that of the observations.}
    \label{fig: model_vs_obs}
\end{figure*}

\begin{figure*}
	\includegraphics[width=\textwidth]{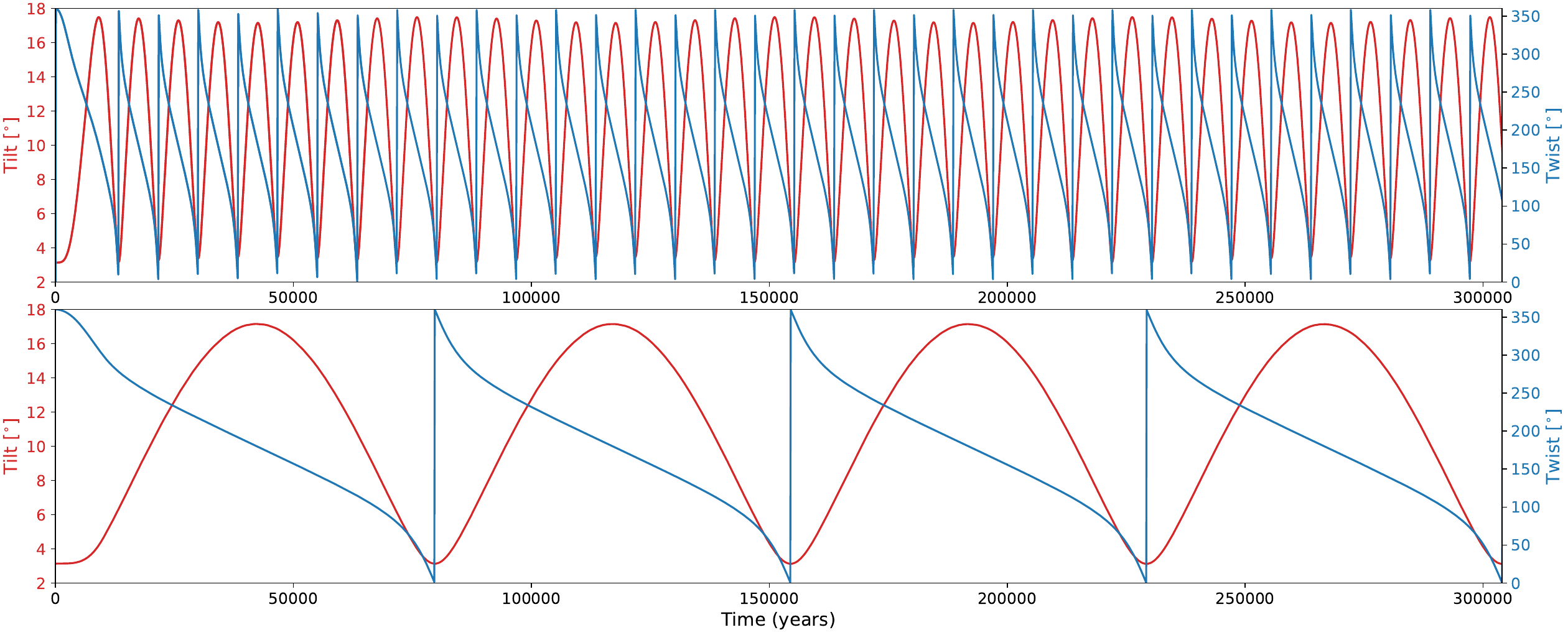}
    \caption{Mean tilt (red) and mean twist (blue) relative to the total angular momentum vector, $\textbf{\textit{l}}_{\rm tot}$, for annuli 1 (top) and 2 (bottom) against time. $\textbf{\textit{l}}_{\rm tot}$ is dominated by the outer disc and so is roughly incident with the $z$-axis. Therefore this plot is useful for understanding the shadows cast on the outer disc (which is in the $xy$-plane). The precession periods of the annuli can be calculated from the twist from this plot as it is measured relative to $\textbf{\textit{l}}_{\rm tot}$ which is conserved. The twist goes from $360^\circ$ to $0^\circ$ as the annuli precess clockwise.}
    \label{fig: twist_tilt_tot_vec}
\end{figure*}

\begin{figure*}
	\includegraphics[width=\textwidth]{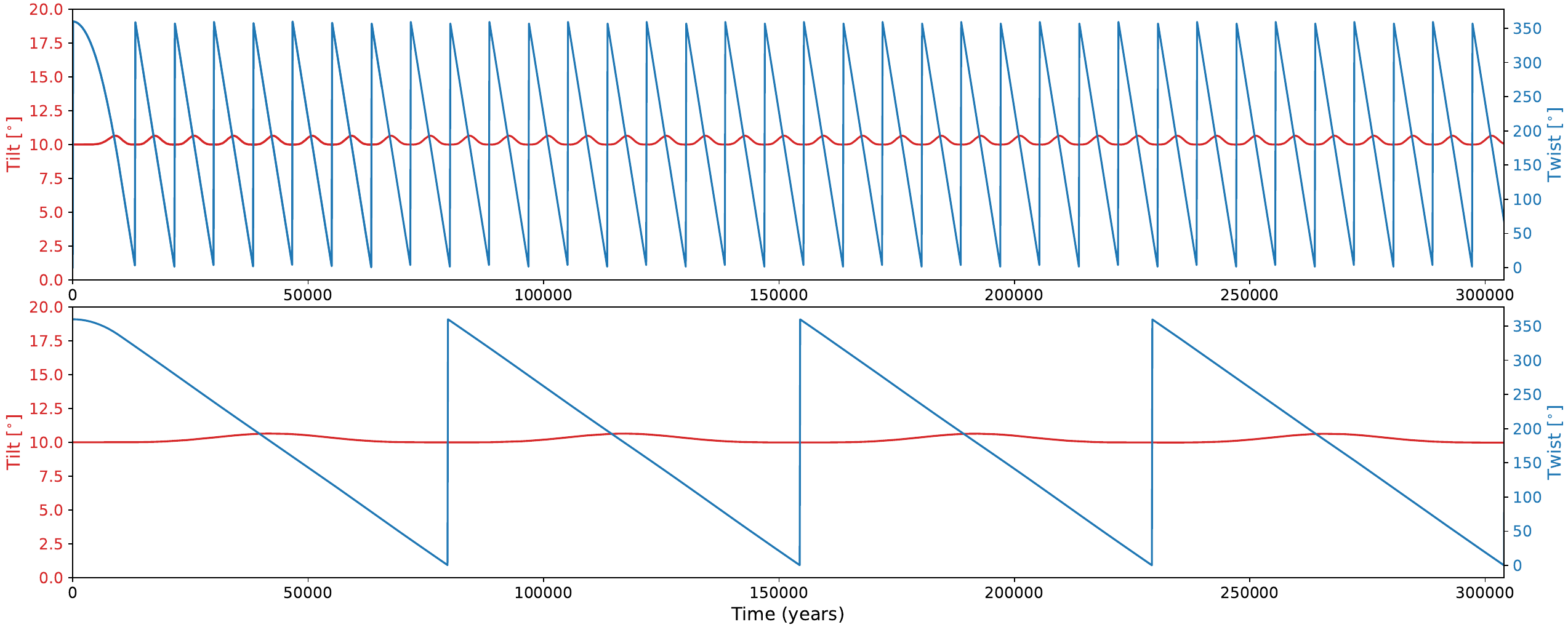}
    \caption{Mean tilt (red) and mean twist (blue) relative to the perturber for annulus 1 (top) and 2 (bottom). This is the same as Figure \ref{fig: twist_tilt_tot_vec}, but revelative to the orbital plane of the perturber, rather than $\textbf{\textit{l}}_{\rm tot}$, so the tilt is zero when the annulus is in the same plane as the perturber's orbit. The annuli precess around the angular momentum vector of the perturber with a constant tilt of $10^\circ$.}
    \label{fig: relative_twist_tilt}
\end{figure*}

The results of our model are plotted in Figure \ref{fig: model_results}. Rows show scattered-light images for different time steps (columns) of the model with a diagram showing the orientation of the inner annuli below each image. This diagram has been inclined by $70^\circ$ (where $0^\circ$ is face-on) and does not show all of annulus 3 (the outer disc). The scattered-light images have been processed in the same way as the observations: they have been scaled with $R^2$ (to counteract the intensity dropping off with the inverse square law) and normalised mean surface brightness of an annulus between $71-134\,$au. The maximum mean from all the images was taken to normalise all the images (for consistency of the colour scale). Images are shown for one precession period of annulus 1.

From Figure \ref{fig: model_results} it is clear that annuli 1 and 2 tilt and precess (with different periods) in a retrograde direction (clockwise) as expected, due to the inclined planet, and the outer disc (annulus 3) remains very flat. Both annuli have relatively low tilts ($< 20^\circ$ - see Figure \ref{fig: twist_tilt_tot_vec}) and so cast a broad shadow across the outer disc. Annulus 1 casts a shadow across both annuli 2 and 3 and itself appears bright, whereas annulus 2 casts a shadow across annulus 3 and is half shadowed by annulus 1. Annulus 2 precesses much more slowly than annulus 1 so the shadow cast by it is consistently present in the North-West and moves very little for the time periods shown in Figure \ref{fig: model_results}, whereas we see the shadow from annulus 1 move round.

Since the annuli precess with different periods the broad shadows cast by each annulus are out of phase and move around at different rates. Therefore, when the shadows are not cast in the same direction but are overlapping (eg. at $\sim2500-3000$ rotations of the planet), they can combine to create a broad shadow across more than half of the outer disc.

Additionally, annulus 2 is only ever shadowed by annulus 1 (unlike the outer disc which is shadowed by both annuli 1 and 2), so only half of it is in shadow at a time which does not necessarily fully align with the shadowing in the outer disc. Therefore, with a coronagraph covering the very inner disc, it appears like the bright ring in HD~139614.

A snapshot of our model can be seen in Figure \ref{fig: model_vs_obs} which reproduces the main features seen in HD~139614: both a broad shadow across two thirds of the outer disc and an inner bright ring. The left of this figure shows a scattered-light image of our model at $47,800\,$years (see Appendix \ref{sec: longer_times} for images of our model over a longer time period) zoomed in to match the scaling of the observations and rotated $165^\circ$ clockwise so that the shadow is in a similar position to the observations. The image is scaled with $R^2$ and normalised (as explained above). A coronagraph has been added in grey to match observations. The right image shows the de-projected, stacked, SPHERE $Q_\phi$ image which has been $R^2$ scaled and normalised (from Figure \ref{fig: stacked_sphere_qphi}) for comparison. 

The model shows a broad shadow across two thirds of the outer disc (annulus 3), which matches that seen in the observations, cast by the combined shadows from annuli 1 and 2. There is also an asymmetric bright ring visible at $13-21\,$au where annulus 2 is partially shadowed due to the shadow cast by the inner annulus (annulus 1). Also, the brightest part does not align with the bright region in the outer disc. Both the broad shadow and bright ring are at angular positions that match that of the observations of HD~139614. 

The inclinations of the annuli in this snapshot are $8.0^\circ$ and $20.0^\circ$ with respect to the $xy-$plane for annulus 1 and 2 respectively with PAs of $351.8^\circ$ and $271.4^\circ$ for the snapshot in Figure \ref{fig: model_vs_obs} (where North is up and East is left). However, we note that these are unlikely to be the only values of inclination and PA which reproduce the observations of HD~139614 (see Section \ref{sec: model_discussion}).

\subsubsection{Periodic precession of the annuli}
\label{sec:precession}
The periodic nature of the precessing annuli can be understood by measuring the mean tilt and twist of the annuli. The tilt, given by:
\begin{equation}
    \arccos{\left(\frac{\textbf{\textit{l}}\cdot\hat{\textbf{\textit{u}}}}{|\textbf{\textit{l}}|}\right)}\, ,
    \label{eq: tilt}
\end{equation}
is the angle between the angular momentum vector of the disc, $\textbf{\textit{l}}$, and a unit vector, $\hat{\textbf{\textit{u}}}$, (such as the $z$-axis) and is a measure of how inclined the disc is with respect to $\hat{\textbf{\textit{u}}}$. The twist measures the rotation of the disc and is calculated as: 
\begin{equation}
    \arctan\left(\frac{l_{\rm y}}{l_{\rm x}}\right)\
    \label{eq: twist}
\end{equation}
where $l_{\rm x}$ and $l_{\rm y}$ are the components of the angular momentum vector of the disc which are perpendicular to $\hat{\textbf{\textit{u}}}$. If $\hat{\textbf{\textit{u}}}=\hat{\textbf{\textit{z}}}$ then $l_{\rm x}$ and $l_{\rm y}$ are the $x$- and $y$-components of the angular momentum vector respectively and the twist is $0^\circ$ in the West (right) and is measured in the West to North direction.

In Figure \ref{fig: twist_tilt_tot_vec} the mean tilt (red) and mean twist (blue) relative to the total angular momentum vector are plotted against time for annuli 1 (top) and 2 (bottom). The total angular momentum vector is calculated from the angular momentum of the disc, perturber, and the star (as it orbits the centre of mass) so $\textbf{\textit{l}}_{\rm tot}=\textbf{\textit{l}}_{\rm disc}+\textbf{\textit{l}}_{\rm planet}+\textbf{\textit{l}}_{\rm star}$ \citep{nealon_2018}. The tilt is then calculated using Equation \ref{eq: tilt} with $\hat{\textbf{\textit{u}}}=\hat{\textbf{\textit{l}}}_{\rm tot}$. We note that the annuli tilt approximately as a rigid body, so that the tilt is the same at every radius in an annulus. This is because the wavelength of the warp is less than the radius of the annulus so the annulus tilts as a whole (eg. \citealt{nixon_pringle_2010}) and globally precesses (eg. \citealt{larwood_1997,fragile_2007,fragner_2010,zhu_2019}). For the twist, the coordinates are transformed so that the $z$-axis is aligned with the total angular momentum vector ($\hat{\textbf{\textit{z}}}'=\hat{\textbf{\textit{l}}}_{\rm tot}$). The twist can then be calculated using Equation \ref{eq: twist} from $l_{\rm x'}$ and $l_{\rm y'}$. 

The total angular momentum is dominated by the outer disc (which is in the $xy$-plane) and so $\textbf{\textit{l}}_{\rm tot}$ is nearly coincident with the $z$-axis. Therefore Figure  \ref{fig: twist_tilt_tot_vec} is useful for understanding the shadows cast on the outer disc. The tilt increases before reaching a peak of $\sim17^\circ$ (\citealt{bitsch_2013} showed that the disc inclination can exceed the planet orbit inclination) then decreases over one precession period. This causes the fluctuation of the shadow depth seen in Figure \ref{fig: model_results}. The twist starts at $360^\circ$ for each annulus and moves down to $0^\circ$ as the annuli precess in a retrograde pattern.

Although Figure \ref{fig: twist_tilt_tot_vec} is useful for understanding the shadows cast on the outer disc, it is not very intuitive for understanding the precession of the annuli. Figure \ref{fig: relative_twist_tilt} shows the twist (blue) and tilt (red) of annuli 1 (top) and 2 (bottom) relative to the angular momentum vector of the perturber, $\textbf{\textit{l}}_{\rm planet}$. The annuli precess around $\textbf{\textit{l}}_{\rm planet}$ with an almost constant tilt of $10^\circ$. The twist has a constant rate of change and moves from $360^\circ$ to $0^\circ$.

The reason the annuli precess around $\textbf{\textit{l}}_{\rm planet}$ is because in this system we are dealing with what are effectively two free low mass annuli (separated by gaps so they behave independently of each other and the outer disc) which are only influenced by the central mass and the perturbing torque of the planet (unlike in \citealt{nealon_2018, zhu_2019} where the disc behaves as a whole). The total angular momentum vector for each annulus is dominated by the planet angular momentum vector (as the mass of each annulus is low) and so they precess around $\textbf{\textit{l}}_{\rm planet}$.

We can calculate the precession periods of the annuli using an equation for the precession of a rigid body (derivation in Appendix \ref{sec:precssion_derivations}) for an annulus with an external perturber (annulus 1):
\begin{align}
    \omega_{\rm{rb}}=-\frac{M_{\rm p}\cos{i_{\rm p}}}{4M_*R_{\rm p}^2}
    \frac{\int^{R_{\rm{out}}}_{R_{\rm{in}}}\Sigma(R)\Omega^2(R)R^5b^{(1)}_{3/2}(\frac{R}{R_{\rm p}})\,dR}{\int^{R_{\rm{out}}}_{R_{\rm{in}}}\Sigma(R)\Omega(R)R^3\,dR}
	\label{eq: precession_inner}
\end{align}
and an annulus with an internal perturber (annulus 2):
\begin{align}
    \omega_{\rm{rb}}=-\frac{M_{\rm p}R_{\rm p}\cos{i_{\rm p}}}{4M_*}
    \frac{\int^{R_{\rm{out}}}_{R_{\rm{in}}}\Sigma(R)\Omega^2(R)R^2b^{(1)}_{3/2}(\frac{R_{\rm p}}{R})\,dR}{\int^{R_{\rm{out}}}_{R_{\rm{in}}}\Sigma(R)\Omega(R)R^3\,dR}
	\label{eq: precession_outer}
\end{align}
where $i_{\rm p}$ is the inclination of the perturber's orbit. These can be solved numerically assuming Keplerian angular frequency and $\Sigma \propto R^{-1}$. Using these equations, we calculate the precession rates for annuli 1 and 2 to be $9.3\times10^3\,$years and $7.3\times10^4\,$years respectively. We calculate the precession periods from our model (from a Fourier transform of the twist in Figures \ref{fig: twist_tilt_tot_vec} and \ref{fig: relative_twist_tilt}) for annulus 1 and 2 to be $8.4\times10^3\,$years and $7.2\times10^4\,$years respectively. The precession periods calculated from Equations \ref{eq: precession_inner} and \ref{eq: precession_outer} are in good agreement with the outcome of our 1D warp model, showing that a rigid body annulus approach is a reasonably reliable approximation to a more sophisticated treatment. Note that approximations that rely on low order expansions of the Laplace coefficients are much less reliable, given the proximity of the annuli to the planet.

\subsection{Model discussion}
\label{sec: model_discussion}

The model is successful in reproducing the gross scattered-light features of HD~139614, specifically the broad shadow and the presence of the bright ring. The model required very little 'tuning'--the annuli are those predicted from SED modelling and the planet radius and mass were taken from predictions from previous studies. The main requirement is to split annulus 2 from the rest of the outer disc, which we refer to as annulus 3. This is done by introducing a gap to stop internal torques exchanging angular momenta between annuli 2 and 3. However gaps are frequently observed in protoplanetary discs \citep{ginski_2016, avenhaus_2018, andrews_2018}, and it does not seem implausible that the disc around HD~139614 has more than one gap (see Section \ref{sec: model_caveats}).

As expected, the precession of the annuli is periodic (see Figure \ref{fig: twist_tilt_tot_vec}) with long precession timescales ($8.4\times10^3$ and $7.2\times10^4\,$years for annuli 1 and 2 respectively). The slow precession rates tell us that the shadows seen in the disc are likely to be long lived and unlikely to change over observable timescales. The shadow cast by the inner annulus would take $\sim23\,$years to move $1^\circ$ which is unlikely to be observable with current instruments (for the shadow cast by annulus 2 a change of $1^\circ$ would take $\sim208\,$years). Therefore the broad shadow seen in HD~139614 could be produced by two precessing annuli and would appear static over the time scales observed. 

However, the slow precession rates of the annuli mean that this model is unlikely to reproduce the temporal variation in the surface brightness of the bright ring which occurs over a $\sim1\,$year time scale (see Section \ref{sec: discussion_obs_bright_ring}). The shadow cast on the bright ring (which is annulus 2) is cast by the inner annulus 1. The period of precession for this annulus is $8.4\times10^3\,$years and so the bright ring would also appear static over very long time scales and certainly over the time scale of our observations. It is likely that the variation in the bright ring is caused by a disc structure close to the host star to have such a short time scale of variation (see Section \ref{sec: discussion_obs_bright_ring} for a discussion on the causes of the temporal variation of the bright ring). 

Additionally, the periodic nature of the precessing annuli means that a broad shadow (across two thirds of the outer disc) occurs multiple times over the disc lifetime and is not just a visible for a specific configuration. Annulus 1 precesses just over 9 times during one precession of annulus 2, creating repeated intervals in which both annuli are sufficiently tilted to cast wide shadows that overlap and cover over half the outer disc. Since the shadow from annulus 2 evolves slowly compared to that of annulus 1, it remains nearly fixed over a single precession period of annulus 1. In contrast, the shadow from annulus 1 sweeps around the disc, periodically aligning with that of annulus 2 to produce a combined shadow extending across approximately two-thirds of the outer disc.

Also, there are multiple occasions where the broad shadow and bright ring have azimuthal positions that match that for HD~139614 (see Figure \ref{fig: model_vs_obs}). For example at $\sim7200$ and $\sim8700$ rotations of the planet (see Appendix \ref{sec: longer_times} for scattered-light images over longer time scales), and at $\sim2900$ and $\sim4300$ rotations of the planet if the disc is viewed from the other side (with an inclination of $180^\circ$ rather than $0^\circ$). This suggests that there are likely multiple configurations of tilted annuli that reproduce the bright ring and broad shadow seen in HD~139614. 

From Figure \ref{fig: twist_tilt_tot_vec} it is clear that the precessing annuli are out of phase as their precession periods differ and are not integer multiples of each other. However, from the beat frequency of these periods we calculate that the configurations of the annuli will repeat every $\sim9\times10^3\,$years. As a result, over multiple periods of precession (of the outer annulus), the shadow configurations will be very similar. Thus the observed broad shadow is not a particularly 'special' epoch of the disc configuration, but rather is predicted to occur over a significant fraction of the precession beat period.

\subsubsection{Comparison with TW Hya}
Our model of tilted, precessing annuli perturbed due to an inclined planet could be applied to other shadowed discs. For instance \citet{debes_2023} propose that the moving shadows in TW Hya could be caused due to two tilted, precessing annuli. In earlier observations of TW Hya there is one precessing shadow \citep{debes_2017}, but in later observations the shadow splits into two which is explained in \citet{debes_2023} by two manually tilted annuli. The annuli extend from $5-6\,$au and $6-7\,$au with moderate inclinations of $5.5^\circ$ and $7^\circ$ (with respect to the outer disc) and PAs of $170^\circ$ and $50^\circ$ respectively. They assume that the annuli precess at Keplerian timescales to reproduce the rotating shadows and do not model the physical motivation for the tilted annuli.

The time period of the precessing shadows is very short: $\sim15\,$yr, so for the precession of the annuli to be driven by a planet on a inclined orbit, the planet is likely to be giant with $M_{\rm p}\sim100\,M_{\rm jup}$ (from analysis of Equations \ref{eq: precession_inner} and \ref{eq: precession_outer}). For both annuli to precess with similar timescales the planet is likely to be in a gap between them at $\sim6\,$au (and the outer annulus must be narrow enough to still have a short precession timescale). The inclination of the planet would likely be $\sim5^\circ-10^\circ$ to reproduce the relatively moderate inclinations of the annuli. Therefore for a precession period of $\sim15\,$yr a $\sim100\,M_{\rm jup}$ planet is required which is much higher than our model for HD~139614 (where $M_{\rm p}=3M_{\rm jup}$). However, a higher planet mass also requires a larger gap width (so that the disc is not within the Hill radius of the planet) which reduces the precession period. Therefore a short precession period on the order of $\sim15\,$yr is unlikely.

We note that \citet{debes_2017} suggest that a planet mass of $\sim19\,M_{\rm jup}$ is enough to perturb an annulus with a precession period of $\sim15\,$yr. However, as pointed out by
\citet{nealon_2019}, there is likely an error in their derivation of the precession period equation (their equation 2) due to unit conversions. \citet{nealon_2019} also suggests that a $12\,M_{\rm jup}$ planet is enough to cause a precession period of $\sim15\,$yr (see their section 4.3), however for this calculation they assume that there is almost no gap between the planet and the edge of the precessing disc, so the disc would be well within the Hill radius of the planet. This configuration is unlikely to be sustained for long due to planet-disc interactions; a gap is required for the planet to maintain its misaligned orbit \citep{nealon_2018} and they conclude that a precessing inner disc is unlikely to explain the moving shadows for TW Hya. Also, both use a less accurate approximation for the precession period than Equation \ref{eq: precession_inner}.

Searches for planets round TW Hya have been unsuccessful to detecting a planet at $\sim6\,$au (NIR data: \citealt{Brandt_2021}; in H$\alpha$: \citealt{cugno_2019,Follette_2023}; and radial velocity data: \citealt{Huelamo_2008}). Measurements of proper motion anomaly (from Hipparcos and Gaia EDR3) limit the mass of potential companions at $1.32$, $1.29$, and $1.30\,M_{\rm Jup}$ for orbits of $3$, $5$, and $10\,$au respectively. This suggests that it is unlikely that a $100\,M_{\rm Jup}$ companion is present at $\sim6\,$au.

Nevertheless, similar configurations, with a combination of rings and a shadow (or shadows) cast due to a misaligned inner disc, can be seen in multiple discs (eg. HD 143006 \citealt{perez_2018, benisty_2018, ballabio_2021,codron_2025}; and Wray 15-788 \citealt{bohn_2019}). Since gaps and rings are thought to be prevalent in protoplanetary discs, tilted, precessing annuli could be a common occurrence.

\subsubsection{Model caveats}
\label{sec: model_caveats}
While our model successfully reproduces the broad shadow and bright ring seen in observations of HD~139614, it does not capture the temporal variation of the bright ring or the spiral arcs (see Section \ref{sec: discussion_spiral_arcs}) as these are beyond the scope of our 1D model. We emphasise that we have produced a physically motivated model which shows a shadow across more than half of the outer disc (which cannot be modelled with just one tilted annulus) and which additionally reproduces the bright ring seen in HD~139614. We do not try to produce a model that replicates all the features from HD~139614. In this section we outline the limitations of our model.

For the annuli in our model to precess independently (and for the model to be stable), there must be gaps between each tilted annulus. The gap between annuli 1 and 2 is supported by interferometric observations of the inner disc and observations of the gas distribution for HD~139614 which both suggest there is a gap at this radius (see Section \ref{sec: intro_hd139614}). In contrast, there is no direct observational evidence of the gap between annuli 2 and 3 (at $21.4-23\,$au). For our model, we include a gap at this location that is sufficiently narrow to be barely visible in our scattered-light models. Therefore it is unlikely to be visible in observations of HD~139614 particularly so close to the coronagraph, where residual starlight increases the noise. However, we note that a gap is required in the gas (so that there is no transfer of angular momentum between annuli), which is less common and harder to detect than gaps in scattered light \citep{van_der_marel_2015, fedele_2017, Dong_2017, van_der_marel_2018}.

From the non-coronagraphic observations in \citet{muro_arena_2020} they see a crescent shaped shadow at $16\,$au which they say may be due to a gap. This suggests that there may be gaps undetected in observations, or that there may be a slightly different configuration of annuli (with a gap at $16\,$au).

Our model also requires a planet on an inclined orbit between annuli 1 and 2: we place a $3M_{\rm jup}$ planet at $4\,\rm au$ inclined by $10^\circ$. There is observational evidence for a planet of this mass at this location (see Section \ref{sec: intro_hd139614}), but the question remains as to how the planet could become inclined. It is possible that the disc became warped, potentially from a perturbing flyby \citep{clarke_pringle_1993, Moeckel_2006, nixon_pringle_2010, milsom_2026}, and the planet then became inclined with respect to the outer disc \citep{xiang_gruess_2016}. Additionally, planets can become inclined due to interactions with each other such as planet-planet scattering \citep{marzari_2002,juri_tremaine_2008,cristobal_2014}, or during planetary migration due to resonant inclination excitation \citep{thommes_lissauer_2003, libert_2009} . However, this requires multiple planets in the system so we prefer the former explanation.

We have chosen reasonable parameters for our model to match HD~139614, but other model parameters may also reproduce the features seen in the observations. For example if we set the planet inclination to $i_{\rm p}=5^\circ$ (see Appendix \ref{sec: beta_5}), which affects how tilted the annuli become, we also reproduce the bright ring and broad shadow. Additionally, scattered-light images are very sensitive to the disc scale height (as mentioned in Section \ref{sec: model setup} where we justify our choice for a flatter disc). The shadows will be affected by the gap width between annuli, and the position and width of the annuli.

\section{Conclusions}
\label{sec: conclusion}
In this paper we have analysed multiple epoch scattered-light images of the protoplanetary disc HD~139614 taken with both GPI and SPHERE instruments. The observations show a broad shadow across two thirds of the outer disc, an azimuthally asymmetric bright ring at $\sim13-21\,$au, and two potentially spiral arcs. We set-up a physically motivated 1D model of the disc coupled with a fast radiative transfer code to produce scattered-light images of our model. Our model reproduces the main features from the observations; namely the broad shadow across two thirds of the outer disc and and inner bright ring.

The main conclusions of our work are summarised below:
\begin{itemize}
    \item We have shown that the azimuthal distribution of the bright ring varies with time. The bright ring appears dimmer in the South (between $\sim125-300^\circ$) in the 2018 epoch than the other epochs (2016, 2017, 2019) and since we see this variation in two independent data sets (from GPI and SPHERE) we strongly suggest that this is a real, physical feature. From the time difference between observations we estimate the time scale of the variation to have a minium of $\sim2$ weeks and a maximum of $\sim11$ months. This short timescale means that the variation is unlikely to be due to a clump of dust moving on Keplerian timescale at the radius of the bright ring ($\sim13-21\,$au), but likely caused by a structure further in (at $\sim1\,$au).
    \item We have shown that the two arcs in the observations may be spiral shaped (as previously suggested by \citealt{laws_2020}) and measure the pitch angle of the spiral for the first time. We fit both Archimedean and logarithmic spirals to both arcs and confirm that a spiral is a good fit for the inner arc (arc 2), but not an improvement on a circular fit for the outer arc (arc 1) with the quality of data we have. We find that the spiral of arc 2 is fairly tightly wound with a pitch angle of $6.1\pm1.2^\circ$.
    \item We produce a physically motivated 1D model of the disc combined with a fast radiative transfer code which reproduces the broad shadow and bright ring seen in scattered-light observations. The model consists of three annuli separated by gaps. The inner two tilt and precess, due to an inclined planet between them, and cast broad shadows on the flat third annulus. At certain time periods these shadows combine to produce a broad shadow across two thirds of the outer disc and the middle annulus appears as a bright ring (half shadowed by the inner most annulus). The brightest part of the bright ring does not align with the brightest part of the outer disc (as in the observations), and both the bright ring and shadow are the same angular positions as the observations.
    \item We show that the two precessing annuli from our model have long precession times ($8.4\times10^3$ and $7.2\times10^4\,$years for annuli 1 and 2 respectively) and are periodic. This tells us that the broad shadow in the outer disc is long-lived (as the annuli precess slowly) and that a broad shadow (across two thirds of the disc) is likely to occur multiple times over a disc's lifetime (as the precession is periodic) and is not just a one off event. Additionally, due to the periodic nature of the precessing annuli, we have shown that the model reproduces the features seen in HD~139614 at multiple time steps suggesting that there may be more than one configuration which reproduces these features.
\end{itemize}

It is important to note that the model presented in the main text is not unique and it not a result of a determined effort to fit the details of the object. We show in Appendix \ref{sec: beta_5} that there are a range of possible planet inclinations that could plausibly reproduce the shadowing effects. Although we focus on one set of parameters this paper, models with different parameters, such as $i_{\rm p}$ or $H/R$, may be able to reproduce these features.

Future work on this system could include repeat observations of HD~139614 in scattered-light to better constrain the timescale of variability of the bright ring. Interferometric and scattered-light observations at matching epochs would allow exploration of how variations in the inner disc structure influence shadowing patterns in the outer disc. Additionally, interferometric observations could also be used to measure the inclination of the inner disc and see if the disc could be made up of tilted annuli.   

High resolution ALMA images would allow us to understand the structure of the midplane of the disc. This could further inform us as to whether the inner disc could be made up of a series of tilted annuli separated by gaps.

Our model of tilted, precessing annuli perturbed due to an inclined planet could be applied to other shadowed discs. For instance \citet{debes_2023} propose that the moving shadows in TW Hya could be caused due to two tilted, precessing annuli (although the time period of the precessing shadows is very short: $\sim15\,$yr). Similar configurations featuring a combination of rings and shadows cast by a misaligned inner disc have been observed in several discs (eg. HD 143006 \citealt{perez_2018, benisty_2018, codron_2025}; and Wray 15-788 \citealt{bohn_2019}). Although we note that for the annuli to tilt and precess independently, gaps are required in the gas, which are less common and harder to detect \citep{van_der_marel_2015, fedele_2017, Dong_2017, van_der_marel_2018}. Since gaps and rings may be a common occurrence in protoplanetary discs, tilted, precessing annuli could be a a widespread phenomenon.

\section*{Acknowledgements}
We thank \citet{muro_arena_2020} for making the SPHERE data available. KLM thanks Daniel Price for useful discussions and for hosting her at Monash University for three months where some of this work was carried out. We thank the anonymous referee for their time and for providing useful comments and feedback that improved the clarity and presentation of this paper. This work was supported by STFC studentship no. 2697199 and the STFC funded long-term attachment. This work made use of NUMPY \citep{harris_2020}, MATPLOTLIB \citep{Hunter_2007}, and SCIPY \citep{virtanen_2020}.
This work is based on observations obtained at the international Gemini Observatory, a program of NSFs NOIR-Lab, which is managed by the Association of Universities for Research in Astronomy (AURA) under a cooperative agreement with the National Science Foundation on behalf of the Gemini Observatory partnership: the National Science Foundation (United States), National Research Council (Canada), Agencia Nacional de Investigación y Desarrollo (Chile), Ministerio de Ciencia, Tecnología e Innovación (Argentina), Ministério da Ciência, Tecnologia, Inovações e Comunicações (Brazil), and Korea Astronomy and Space Science Institute (Republic of Korea).
SPHERE is an instrument designed and built by a consortium consisting of IPAG (Grenoble, France), MPIA (Heidelberg, Germany), LAM (Marseille, France), LESIA (Paris, France), Laboratoire Lagrange (Nice, France), INAF – Osservatorio di Padova (Italy), Observatoire de Genève (Switzerland), ETH Zurich (Switzerland), NOVA (The Netherlands), ONERA (France), and ASTRON (The Netherlands) in collaboration with ESO. SPHERE was funded by ESO, with additional contributions from CNRS (France), MPIA (Germany), INAF (Italy), FINES (Switzerland), and NOVA (The Netherlands). SPHERE also received funding from the European Commission Sixth and Seventh Framework Programmes as part of the Optical Infrared Coordination Network for Astronomy (OPTICON) under grant number RII3-
Ct2004-001566 for FP6 (2004-2008), grant number 226604 for FP7 (2009-2012), and grant number 312430 for FP7 (2013-2016).
ASEL notes that this report is independent research supported by the National Institute for Health and Care Research Applied Research Collaboration South West. The views expressed in this publication are those of the author(s) and not necessarily those of the National Institute for Health Research or the Department of Health and Social Care.
\section*{Data Availability}
The GPI data used in this work are available from \citet{rich_2022_data} and the SPHERE data from \citet{muro_arena_2020}. The 1D warp modelling code used to carry out this work is available via Zenodo (\url{https://doi.org/10.5281/zenodo.20557882}). 

\bibliographystyle{mnras}
\bibliography{bib} 



\appendix

\section{\texorpdfstring{$U_{\phi}$}{}}
\label{sec: u_phi}

\begin{figure*}
	\includegraphics[width=\textwidth]{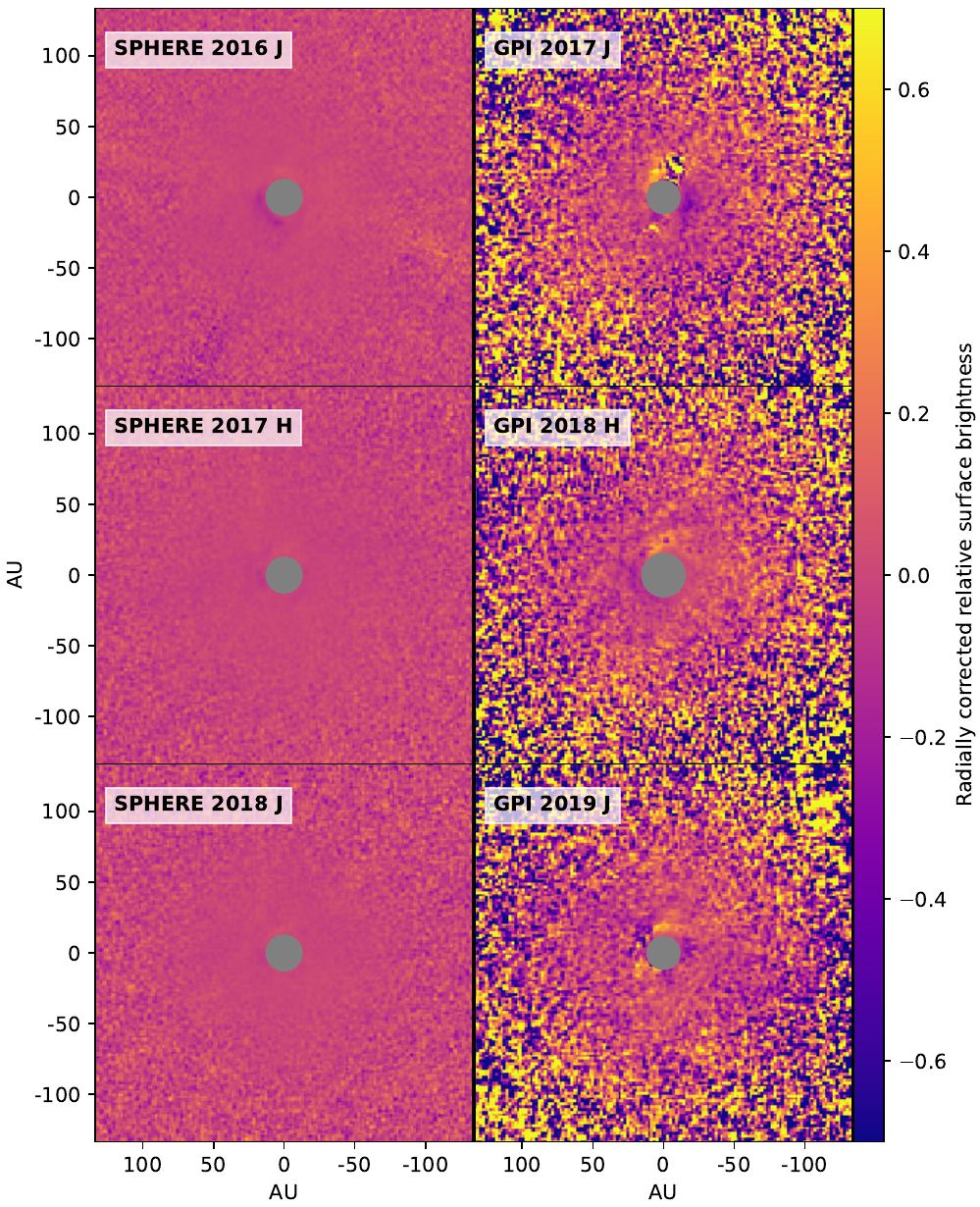}
    \caption{$U_{\phi}$ images of the de-projected SPHERE (left) and GPI (right) data for HD139614. The data has been $R^2$ scaled and normalised to the $95^{\rm th}$ percentile of the surface brightness. The labels show the instrument, epoch, and wavelength band for each image. The coronagraphic spot is marked by a grey circle. North is up and East is left for all images.}
    \label{fig: uphi_data_full_plot}
\end{figure*}

Figure \ref{fig: uphi_data_full_plot} shows the $U_{\phi}$ data for HD~139614. The images have been de-projected, scaled and normalised to the $95^{\rm th}$ percentile of the surface brightness (for the brightest image). The images are mostly featureless suggesting that there is little to no self-scattering occurring and that there are no companions visible. These images show that the noise is larger for the GPI images as there are many bright pixels in the image (especially at larger radii where the noise is emphasised by the scaling). A bright spot can be seen in the North of the GPI 2017 epoch near the coronagraph which lines up with the dark spot in the $Q_{\phi}$ data. This may be due to the telescope pixels becoming oversaturated.

\section{Derivation of precession equations}
\label{sec:precssion_derivations}
Here we derive Equations \ref{eq: precession_inner} and \ref{eq: precession_outer} for the rigid body precession of annuli 1 and 2. The equation for the angular precession of an annulus of infinitesimal width, $dR$, is given by \citep{murray_dermott_1999}:
\begin{equation}
    \omega_{\rm p}=-\frac{1}{4}\frac{M_{\rm p}}{M_*}\Omega(R) \alpha\bar{\alpha}  b^{(1)}_{3/2}(\alpha)\cos{i_{\rm p}}
    \label{eq: wp_general}
\end{equation}
where the values of $\alpha$ and $\bar{\alpha}$ depend on whether the orbit of perturbing companion is internal or external to the annulus given by:
\begin{align}
    \alpha=
    \begin{cases}
    \frac{R_{\rm p}}{R}, & \text{if } R > R_{\rm p} \\
    \frac{R}{R_{\rm p}}, & \text{if } R < R_{\rm p}
    \end{cases}
    \label{eq: alpha}
\end{align}

\begin{align}
    \bar{\alpha}=
    \begin{cases}
    1, & \text{if } R > R_{\rm p} \\
    \frac{R}{R_{\rm p}}, & \text{if } R < R_{\rm p}\ .
    \end{cases}
    \label{eq: alpha_bar}
\end{align}
Therefore, for an annulus (of width $\rm dR$) with an internal perturber ($R>R_{\rm p}$) the angular precession is given by:
\begin{equation}
    \omega_{\rm p}=-\frac{1}{4}\frac{M_{\rm p}}{M_*}\Omega(R)\left(\frac{R_{\rm p}}{R}\right)b^{(1)}_{3/2}\left(\frac{R_{\rm p}}{R}\right)\cos{i_{\rm p}}
    \label{eq: wp_inner}
\end{equation}
and for an annulus with an external perturber ($R<R_{\rm p}$):
\begin{equation}
    \omega_{\rm p}=-\frac{1}{4}\frac{M_{\rm p}}{M_*}\Omega(R)\left(\frac{R}{R_{\rm p}}\right)^2b^{(1)}_{3/2}\left(\frac{R}{R_{\rm p}}\right)\cos{i_{\rm p}}\ .
    \label{eq: wp_outer}
\end{equation}

However, we wish to know the precession rate of a full annulus, not just with infinitesimal width $dR$. We can assume that the annulus precesses as a rigid body since there is communication throughout the disc. The precession of a rigid body is given by:
\begin{equation}
    \omega_{\rm rb}=\frac{\int^{R_{\rm out}}_{R_{\rm in}}\omega_{\rm p}\,dL}{\int^{R_{\rm out}}_{R_{\rm in}}\,dL}
    \label{eq: rigid_body_integral}
\end{equation}
where $dL$ is the angular momentum of an annulus at radius $R$ with width $dR$ given by:
\begin{equation}
    \mathrm{dL}=\Sigma(R)\Omega(R)R^3\,\mathrm{d}R\ .
\end{equation}
Substituting this into Equation \ref{eq: rigid_body_integral} gives:
\begin{align}
    \omega_{\rm{rb}}=\frac{\int^{R_{\rm{out}}}_{R_{\rm{in}}}\Sigma(R)\omega_{\rm p}\Omega(R)R^3\,dR}{\int^{R_{\rm{out}}}_{R_{\rm{in}}}\Sigma(R)\Omega(R)R^3\,dR}\ .
    \label{eq: wrb}
\end{align}

Therefore, for the angular precession of an annulus with an internal perturber we substitute Equation \ref{eq: wp_outer} into Equation \ref{eq: wrb} to get Equation \ref{eq: precession_outer} :
\begin{align}
    \omega_{\rm{rb}}=-\frac{M_{\rm p}R_{\rm p}\cos{i_{\rm p}}}{4M_*}
    \frac{\int^{R_{\rm{out}}}_{R_{\rm{in}}}\Sigma(R)\Omega^2(R)R^2b^{(1)}_{3/2}(\frac{R_{\rm p}}{R})\,dR}{\int^{R_{\rm{out}}}_{R_{\rm{in}}}\Sigma(R)\Omega(R)R^3\,dR}\ .
\end{align}
For an annulus with an external perturber we substitute Equation \ref{eq: wp_inner} into Equation \ref{eq: wrb} to get Equation \ref{eq: precession_inner}:
\begin{align}
    \omega_{\rm{rb}}=-\frac{M_{\rm p}\cos{i_{\rm p}}}{4M_*R_{\rm p}^2}
    \frac{\int^{R_{\rm{out}}}_{R_{\rm{in}}}\Sigma(R)\Omega^2(R)R^5b^{(1)}_{3/2}(\frac{R}{R_{\rm p}})\,dR}{\int^{R_{\rm{out}}}_{R_{\rm{in}}}\Sigma(R)\Omega(R)R^3\,dR}\ .
\end{align}

\section{scattered-light images for a longer time period}
\label{sec: longer_times}

\begin{figure*}
\includegraphics[width=0.90\textwidth]{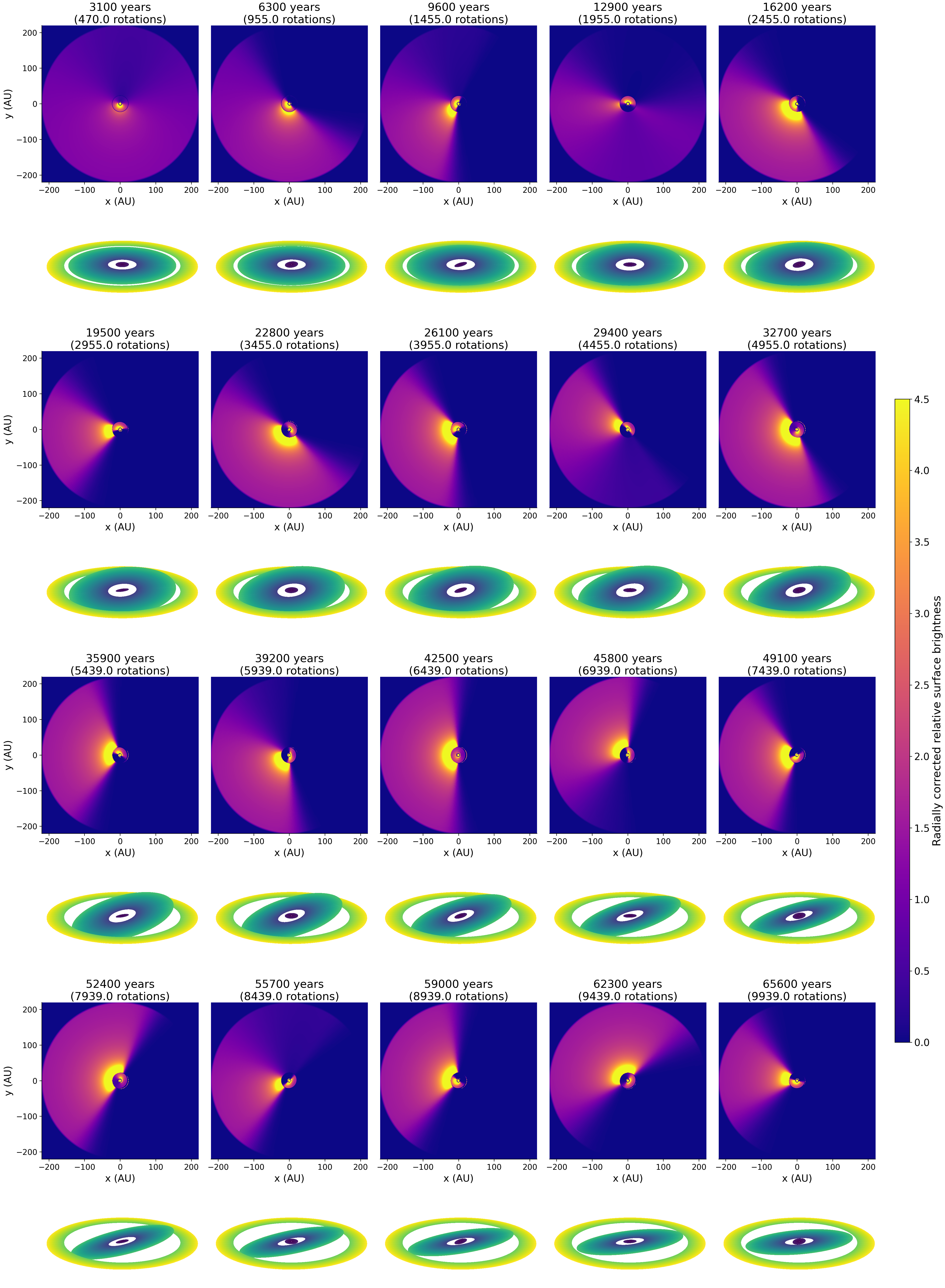}
    \caption{The same as Figure \ref{fig: model_results}, but for images over a longer time period ($\sim 10,000\,$rotations of the planet rather than just $\sim 3500$). Rows show scattered-light images at different time steps (columns) with a diagram below each image of the inner disc. The diagrams are inclined by $70^\circ$ (where $0^\circ$ is face-on) and not all of annulus 3 is shown. The scattered-light images have been scaled by $R^2$ and normalised. The shadow due to annulus 2 precesses clockwise so it starts in the North-West ($\sim 900\,$rotations) and rotates to the South-West by $\sim 10,000\,$rotations. The shadow due to annulus 1 precesses on a much quicker timescale so we only see snapshots of it's movement.}
    \label{fig: model_longer_time_steps}
\end{figure*}

Figure \ref{fig: model_longer_time_steps} shows scattered-light images of our model for $\sim10,000\,$rotations of the planet. It is the same as Figure \ref{fig: model_results}, but over a time period $\sim3$ times as long. Scattered-light images (rows) are shown for different time steps (columns) with a diagram of the disc (which has been inclined by $70^\circ$ where $0^\circ$ is face-on) below each image. Not all of annulus 3 is shown in the diagrams. The scattered-light images have been scaled with $R^2$ and normalised (as explained in Section \ref{sec: model_results}).

We can see the broad shadow cast by the slowly precessing annulus 2 rotate clockwise from the North-West (at $\sim900\,$rotations) to the South-West at $\sim10,000\,$rotations. Annulus 1 is precessing too fast to track the shadow rotation over this time period so we just see snapshots of its rotation (Figure \ref{fig: model_results} shows the precession of the shadow due to annulus 1).

\section{scattered-light images for model with a planet inclination of \texorpdfstring{$5^\circ$}{} }
\label{sec: beta_5}

\begin{figure*}
\includegraphics[width=0.91\textwidth]{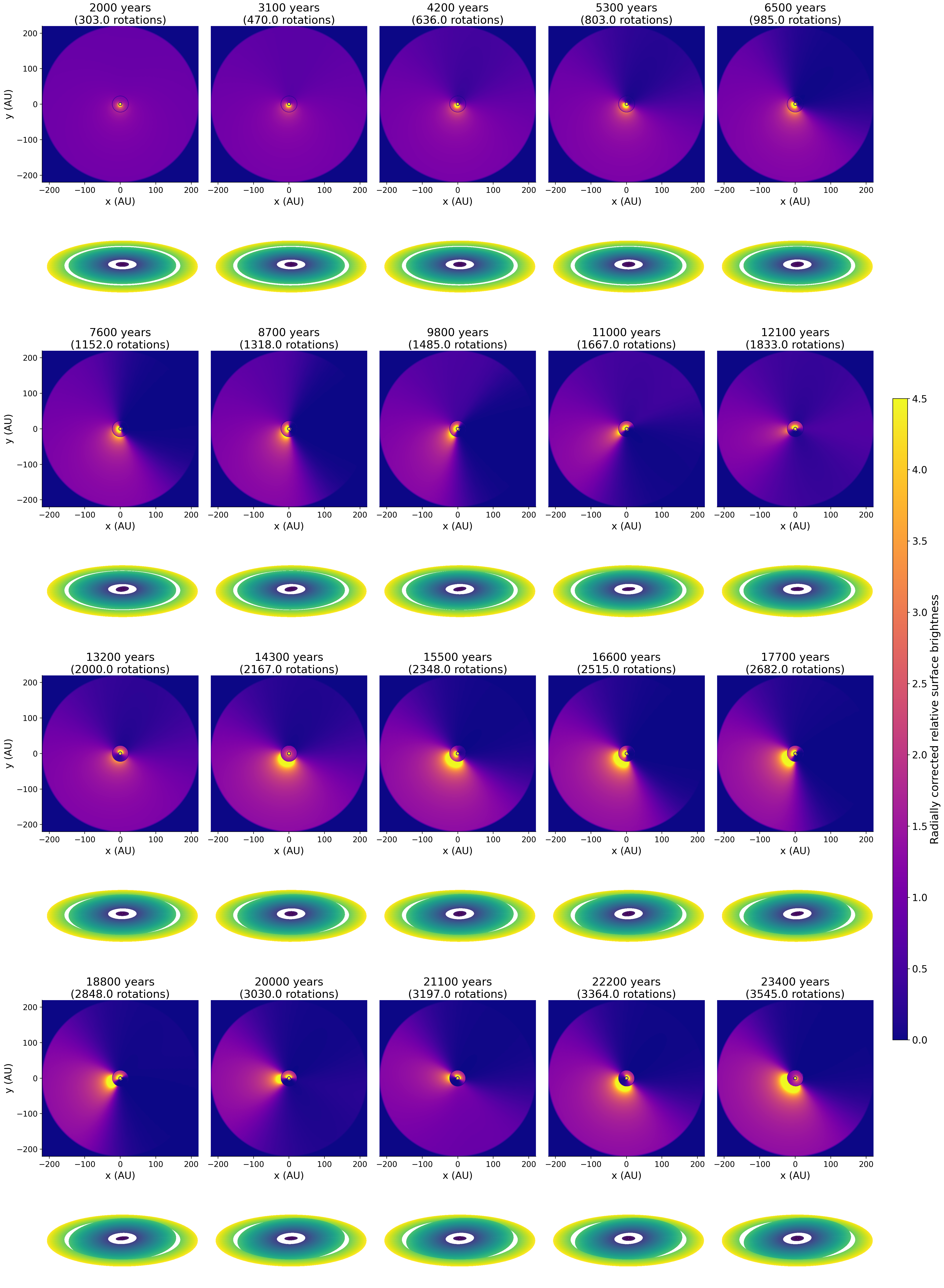}
    \caption{The same as Figure \ref{fig: model_results} but the planet is inclined by $5^\circ$ rather than $10^\circ$. Rows show scattered-light images at different time steps (columns) with a diagram below each image of the inner disc. The diagrams are inclined by $70^\circ$ (where $0^\circ$ is face-on) and not all of annulus 3 is shown. The scattered-light images have been scaled by $R^2$ and normalised. The shadows are very similar to that in Figure \ref{fig: model_results}, but the relative surface brightness is fainter and so the shadows are more subtle.}
    \label{fig: model_beta5}
\end{figure*}

Here we present the results for our model where the planet is inclined by $5^\circ$ ($i_{\rm p}=5^\circ$) rather than $10^\circ$ (as in the main body of the paper). Figure \ref{fig: model_beta5} shows the scattered-light images from our model (rows) for different time periods (columns). This figure is the same as Figure \ref{fig: model_results}, but for $i_{\rm p}=5^\circ$. A diagram of the inner disc is shown below each image. The disc diagram is inclined by $70^\circ$ (where $0^\circ$ is face-on) and not all of annulus 3 is shown. The scattered-light images have been scaled with $R^2$ and normalised as explained in Section \ref{sec: model_results}.

The scattered-light images show the same broad shadows (which combine to produce a shadow over more than half the disc between $\sim17,000-20,000\,$years) as Figure \ref{fig: model_results}, but the shadows are dimmer (the relative surface brightness is lower) and so more subtle. This is because although for $i_{\rm p}=5^\circ$ the annuli have the same precession periods as for $i_{\rm p}=10^\circ$, they are less tilted; their maximum tilt is $10^\circ$ rather than $20^\circ$ and so cast dimmer shadows. Therefore we focus on the model with $i_{\rm p}=10^\circ$ in the main body of this paper as the shadows are clearer.

A key take away from these results is that not just one model can reproduce the broad shadow and bright ring seen in observations of HD~139614. Our model with $i_{\rm p}=5^\circ$ also reproduces the broad shadow across $\sim2/3$ of the disc and the bright ring for the equivalent snapshot as that shown in Figure \ref{fig: model_vs_obs}. Although we focus on one set of parameters in the main body of the paper, models with different parameters (such as $i_{\rm p}$) may be able to reproduce these features.


\bsp	
\label{lastpage}
\end{document}